\documentclass[aps,prx,twocolumn,superscriptaddress,longbibliography]{revtex4-1}
\usepackage{float}
\usepackage{graphicx}
\usepackage{dsfont}
\usepackage[latin1]{inputenc}
\usepackage{version} 
\usepackage{color}
\usepackage{amssymb}
\usepackage{amsmath}
\usepackage{braket}
\usepackage{bm}
\usepackage{upgreek}
\usepackage{colortbl}
\usepackage{graphicx}
\usepackage{dcolumn}
\usepackage{bm}
\usepackage{bigints}
\usepackage{xcolor}  
\usepackage[pagebackref=false]{hyperref} 
\usepackage{enumitem}

\setcitestyle{super}

\definecolor{jrp}{rgb}{1,0,0}

\definecolor{rim}{rgb}{0,1,0}

\definecolor{pr}{rgb}{0.7,0,0}

\definecolor{mjg}{rgb}{.08,.05,.8}

\definecolor{yyl}{rgb}{.8,.05,.08}

\newcommand{\delete}[1]{{}} 
\colorlet{shadecolor}{gray!40}

\hypersetup{
    bookmarks=true,         
    unicode=false,          
    pdftoolbar=true,        
    pdfmenubar=true,        
    pdffitwindow=false,     
    pdfstartview={FitH},    
    pdftitle={DTC_Main},    
    pdfauthor={X. Mi},     
    pdfkeywords={keyword1} {key2} {key3}, 
    pdfnewwindow=true,      
    colorlinks=true,       
    linkcolor=red,          
    citecolor=blue,        
    filecolor=magenta,      
    urlcolor=cyan,           
}

\makeatletter

\newcommand{\avA}{\ensuremath{\overline{A}}} 
\newcommand{\avnorm}{\ensuremath{\overline{A}_0}} 
\newcommand{\avavA}{\ensuremath{[\overline{A}]}} 

\makeindex

\begin{document}
\title{Observation of Time-Crystalline Eigenstate Order on a Quantum Processor}
\author{Google Quantum AI and collaborators$^{\hyperlink{authorlist}{\dagger},}$}

\email[Corresponding authors: V.~Khemani: ]{vkhemani@stanford.edu}
\email[\\ P. Roushan: ]{pedramr@google.com}

\maketitle

{\bf 
Quantum many-body systems display rich phase structure in their low-temperature equilibrium states\cite{Wentextbook}. However, much of nature is not in thermal equilibrium. Remarkably, it was recently predicted that out-of-equilibrium systems can exhibit novel dynamical phases\cite{ Bukov_2015, HarperReview2020, Oka_2009, Rudner2013,AFAI,Khemani2016, Po2016} that may otherwise be forbidden by equilibrium thermodynamics, a paradigmatic example being the discrete time crystal (DTC)\cite{Khemani2016, Else2016, vonKeyserlingk2016, Sacha_2017, Khemani2019, Bruno, Watanabe_2015}. Concretely, dynamical phases can be defined in periodically driven many-body localized systems via the concept of eigenstate order\cite{Khemani2016, Huse2013lpqo, PekkerHilbertGlass}. In eigenstate-ordered phases, the \emph{entire} many-body spectrum exhibits quantum correlations and long-range order, with characteristic signatures in late-time dynamics from \emph{all} initial states. It is, however, challenging to experimentally distinguish such stable phases from transient phenomena, wherein few select states can mask typical behavior. Here we implement a continuous family of tunable CPHASE gates on an array of superconducting qubits to experimentally observe an eigenstate-ordered DTC. We demonstrate the characteristic spatiotemporal response of a DTC for generic initial states\,\cite{Khemani2016, Else2016, vonKeyserlingk2016}. Our work employs a time-reversal protocol that discriminates external decoherence from intrinsic thermalization, and leverages quantum typicality to circumvent the exponential cost of densely sampling the eigenspectrum. In addition, we locate the phase transition out of the DTC with an experimental finite-size analysis. These results establish a scalable approach to study non-equilibrium phases of matter on current quantum processors.}

\begin{figure}[t!]
	\centering
	\includegraphics[width=1\columnwidth]{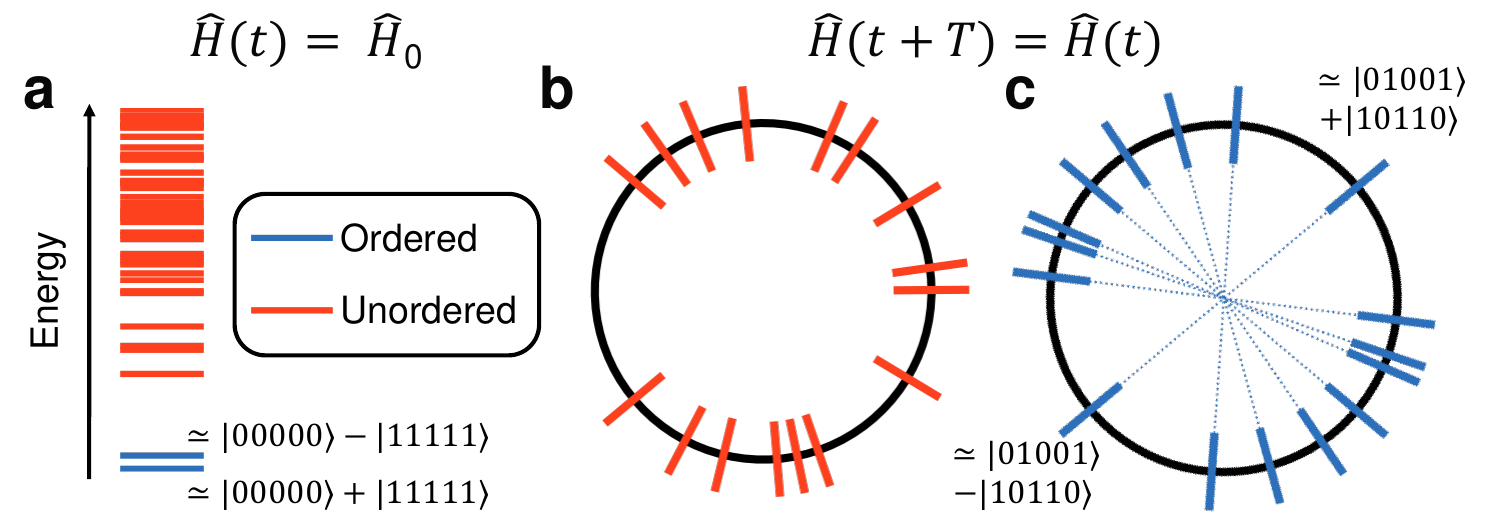}
	\caption{\textbf{Order in eigenstates.} {\bf a}, Equilibrium phases are characterized by long-range order in low-energy eigenstates of time-independent Hamiltonians, e.g. an Ising ferromagnet with a pair of degenerate ground states that resemble ``Schr\"odinger cats'' of polarized states. {\bf b}, Floquet systems typically have no ordered states in the spectrum. {\bf c}, In MBL Floquet systems, every eigenstate can show order. In MBL-DTC, every eigenstate resembles a long-range ordered ``Schr\"odinger cat'' of a random configuration of spins and its inversion, with even/odd superpositions split by $\pi$.
	}
	\label{fig:1}
\end{figure}

In an equilibrium setting, quantum phases of matter are classified by long-range order or broken symmetries in low-temperature states (Fig.~\ref{fig:1}a). The existence of ordered phases in periodically driven (Floquet) systems, on the other hand, is counterintuitive: Since energy is not conserved, one expects thermalization to a featureless maximum-entropy state that is incompatible with quantum order. However, this heat death is averted in the presence of many-body localization (MBL), where strong disorder causes the emergence of an extensive number of local conservation laws which prevent thermalization\cite{Basko2006, Nandkishore2015, AbaninRMP2019,  Ponte_2015, Lazarides_2015, Bordia_2017}, making it possible to stabilize intrinsically dynamical phases\cite{Khemani2016}. 

Dynamics in a Floquet system is governed by a unitary time evolution operator, whose eigenvalues lie on the unit circle. While the entire Floquet spectrum is featureless in a thermalizing phase (Fig.~\ref{fig:1}b), an MBL Floquet phase can have an order parameter associated with each eigenstate. As an example, in the spatiotemporally-ordered MBL-DTC, the spectrum has a distinctive pattern of pairing between ``Schr\"odinger cat'' states that are separated by an angle $\pi$ (Fig.~\ref{fig:1}c)\cite{Khemani2016, Else2016, vonKeyserlingk2016}. This pairing manifests as a stable sub-harmonic response, wherein local observables show period-doubled oscillations that spontaneously break the discrete time translation symmetry of the drive for infinitely long times. The unique combination of spatial long-range order and time translation symmetry breaking in an isolated dissipation-free quantum many-body system is the hallmark of the MBL-DTC.

Experimentally observing a non-equilibrium phase such as the MBL-DTC is a challenge due to limited programmability, coherence and size of Noisy Intermediate Scale Quantum (NISQ) hardware. 
Sub-harmonic response, by itself, is not a unique attribute of the MBL-DTC; rather, it is a feature of many dynamical phenomena whose study has a rich history\cite{faradaywaves, Goldstein2018} (See also Ch.~8 in Ref.~\onlinecite{Khemani2019}). Most recently, interesting DTC-like dynamical signatures have been observed in a range of quantum platforms\cite{LukinExp, MonroeExp, BarrettExp, Sreejith}. Such signatures, however, are transient, arising from slow or prethermal dynamics from special initial states\cite{Else2017, Ho2017, Luitz2020, Khemani2019, Ippoliti2020}, and are separated from the MBL-DTC by a spectral phase transition where eigenstate order disappears. Thus, despite the recent progress, observing an MBL-DTC remains an outstanding challenge\cite{Khemani2019, Ippoliti2020}. 

\begin{figure*}[t!]
	\centering
	\includegraphics[width=2\columnwidth]{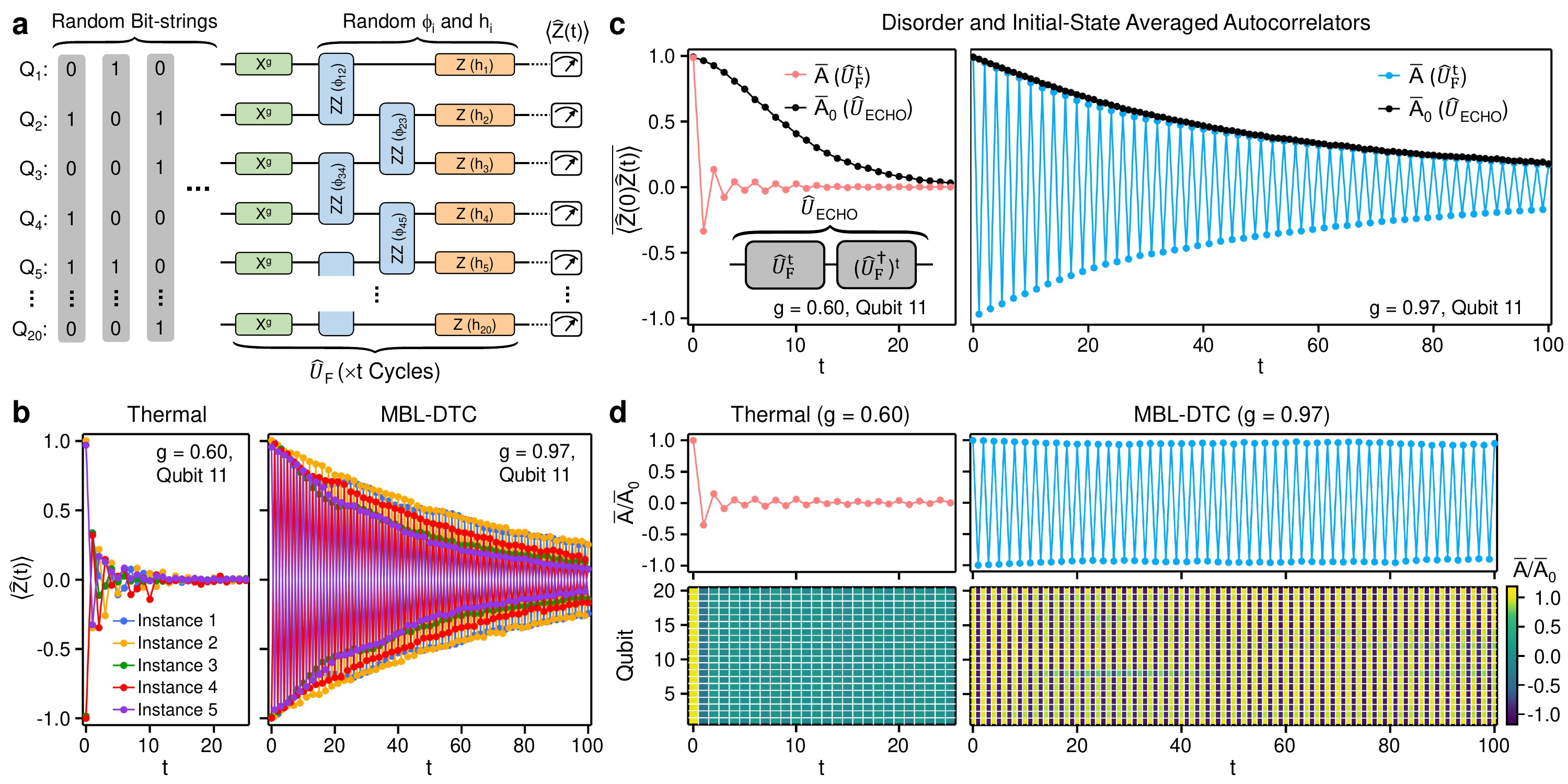}
	\caption{\textbf{Observing a many-body localized discrete time-crystal.} \textbf{a,} Schematic of the experimental circuit composed of $t$ identical cycles of the unitary $\hat{U}_\text{F}$. The local polarization of each qubit, $\braket{\hat{Z} (t)}$, is measured at the end. In the following panels, we investigate a number of disorder instances each with a different random bit-string initial state. \textbf{b,} Experimental values of $\braket{\hat{Z} (t)}$ measured at qubit 11. Data are shown for five representative circuit instances deep in the thermal ($g = 0.60$) and MBL-DTC ($g = 0.97$) phases. \textbf{c,} Autocorrelator $\avA = \overline{\braket{\hat{Z} (0) \hat{Z} (t)}}$ at qubit 11, obtained from averaging the results of 36 circuit instances. For the same circuit instances, the average autocorrelator at the output of $\hat{U}_\text{ECHO} = (\hat{U}_\text{F}^\dagger)^t \hat{U}^t_\text{F}$ is also measured and its square root, $\avnorm$, is shown alongside $\avA$ for comparison. \textbf{d,} Top panels: The ratio $\avA / \avnorm$ obtained from panel \textbf{c}. Bottom panels: $\avA / \avnorm$ as a function of $t$ and qubit location. 
}\label{fig:2}
\end{figure*} 

Here we perform the following necessary benchmarks for experimentally establishing an eigenstate-ordered non-equilibrium phase of matter: (i) Drive parameters are varied in order to demonstrate stability of the phase in an extended parameter region and across disorder realizations; The limitations of (ii) finite size and (iii) finite coherence time are addressed, respectively, by varying the number of qubits in the system and by separating effects of extrinsic decoherence from intrinsic thermalization; (iv) The existence of eigenstate order across the {\it entire} spectrum is established. The flexibility of our quantum processor, combined with the scalable experimental protocols devised in the following, allows us to fulfill these criteria and observe time-crystalline eigenstate order. 

The experiment is conducted on an open-ended, linear chain of $L=20$ superconducting transmon qubits ($Q_1$ through $Q_{20}$) that are isolated from a two-dimensional grid. We drive the qubits via a time-periodic (Floquet) circuit $\hat{U}_\text{F}^t$ with $t$ identical cycles (Fig.~\ref{fig:2}a) of $\hat{U}_\text{F}$: 
\begin{equation}
\hat{U}_F = 
\underbrace{e^{-\frac{i}{2} \sum_i h_i \hat{Z}_i} }_\text{longitudinal fields}
\underbrace{e^{-\frac{i}{4} \sum_i \phi_i \hat{Z}_i \hat{Z}_{i+1}} }_\text{Ising interaction}
\underbrace{e^{-\frac{i}{2} \pi g \sum_i  \hat{X}_i}}_{x \text{ rotation by } \pi g}
\label{eq:model}
\end{equation}
where $\hat{X}_i$ and $\hat{Z}_i$ are Pauli operators. Each $\phi_i$ ($h_i$) is sampled randomly 
from $[-1.5 \pi, -0.5 \pi]$ ($[-\pi,\pi]$) for every realization of the circuit.  Overall, $\hat{U}_\text{F}$ implements an interacting Ising model that is periodically ``kicked'' by a transverse pulse that rotates all qubits by $\pi g$ about the $x$ axis. In this work, $g$ is tuned within the range $[0.5, 1.0]$ to explore the DTC phase and its transition into a thermal phase. At $g=1$, the model implements a $\pi$ pulse which exactly flips all qubits (in the $z$ basis) and returns them to the initial state over two periods. A key signature of the DTC is the presence of \emph{robust} period doubling, \emph{i.e.} extending over a finite extent in parameter space, even as $g$ is tuned away from $1$. Strong Ising interactions, which produce long-range \emph{spatial} order, are essential for this robustness\cite{Khemani2016, vonKeyserlingk2016}. This is in contrast to a system of decoupled qubits ($\phi=0$) which rotate by a continuously varying angle $\pi g$ every period instead of being locked at period doubling. Prior theoretical work\cite{Ippoliti2020} has shown that model~\eqref{eq:model} is expected to be in an MBL DTC phase in the range $g > g_\text{c}$, and transition to a thermal phase at a critical value\ $g_\text{c} \approx 0.84$.

\begin{figure*}[t!]
	\centering
	\includegraphics[width=2\columnwidth]{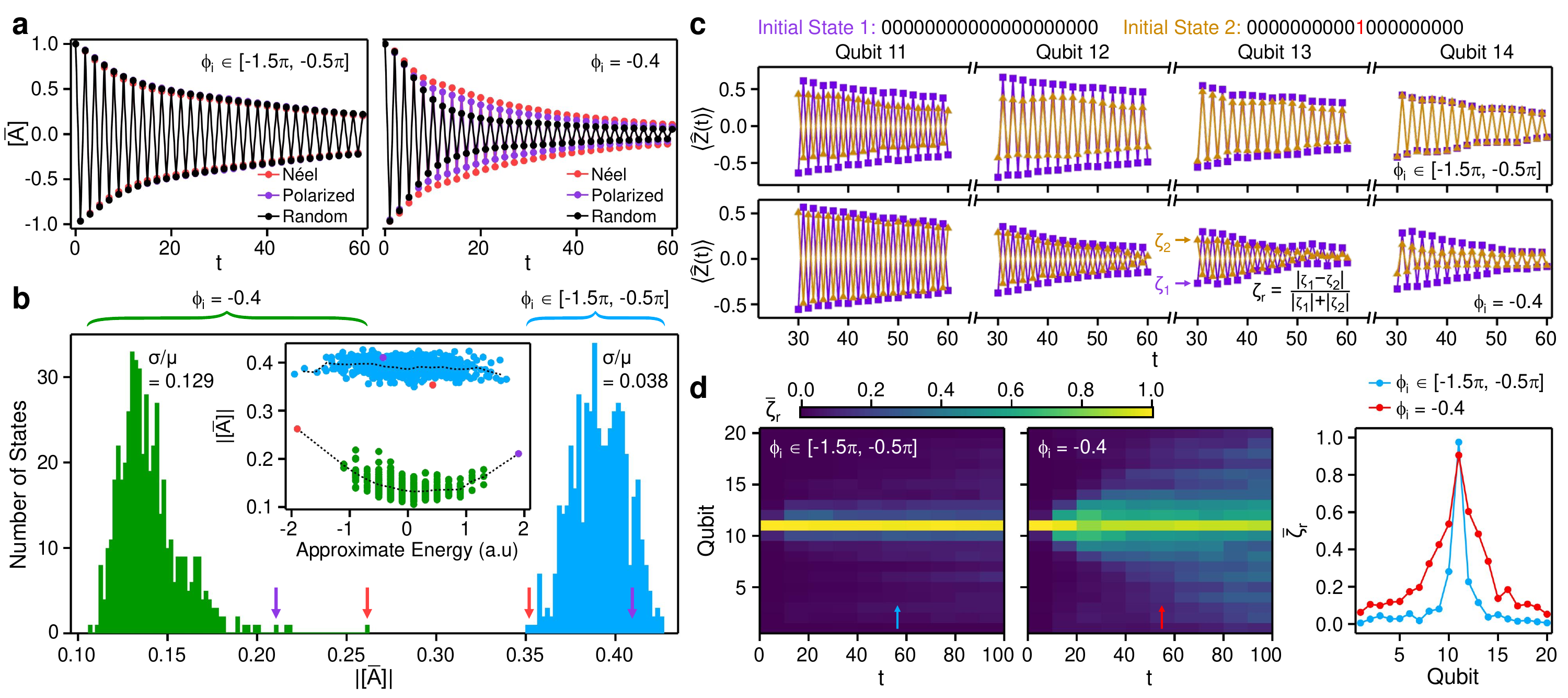}
	\caption{\textbf{Observing eigenstate order and distinguishing it from transient phenomena.} \textbf{a,} Site- and disorder-averaged autocorrelators $\avavA$ measured with $g = 0.94$. In the left panel (MBL-DTC), each data set is averaged over 24 disorder instances of $\phi_i$ and $h_i$, with the initial state fixed at one of the following: N\'eel: $\ket{01}^{\otimes 10}$, Polarized: $\ket{0}^{\otimes 20}$, Random: $\ket{00111000010011001111}$. In the right panel (thermalizing), the same values of $h_i$ and initial states are used but $\phi_i=-0.4$. \textbf{b,} Histograms of $|\avavA|$, from 500 random bit-string initial states, averaged over cycles 30 and 31 and the same disorder instances as in panel \textbf{a}. $\sigma / \mu$, where $\sigma$ ($\mu$) is the standard deviation (mean) of $|\avavA|$, is also listed. Location of the polarized (N\'eel) state is indicated by a purple (red) arrow. Inset: same collection of $|\avavA|$ plotted over the energies of the bit-string states, calculated from the effective Hamiltonian $\hat{H}_\text{eff}$ approximating the drive (see text). Dashed lines show averaged values within energy windows separated by 0.2.
	\textbf{c,} $\braket{\hat{Z} (t)}$ for two bit-string initial states that differ only at $Q_{11}$. Top panel shows a single circuit instance with disordered $\phi_i$  
	and bottom panel shows an instance with uniform $\phi_i = -0.4$. \textbf{d,} Left and middle panels: Relative difference between the two signals $\overline{\zeta}_\text{r}$ as a function of $t$ and qubit location, averaged over time windows of 10 cycles and over 64 disorder instances for $\hat{U}_F$ and 81 instances for $\hat{U}_F'$. Right panel: Qubit dependence of $\overline{\zeta}_\text{r}$, averaged from $t = 51$ to $t = 60$.
}
	\label{fig:3}
\end{figure*}

Achieving MBL in this model for $g \sim 1$ requires disorder in the two-qubit interaction, $\phi_i$, which is even under Ising symmetry\cite{Khemani2019, Ippoliti2020}, $\prod_i \hat{X}_i$, a condition that was not met by some past DTC experiments~\cite{MonroeExp, BarrettExp}. Ising-odd terms, {\it i.e.} $h_i$, are approximately dynamically decoupled by the $x$ pulses over two periods, thereby lowering their effective disorder strength and hindering localization (in the absence of independent disorder in the $\phi_i$). Utilizing newly developed CPHASE gates (see SM for details) with continuously tunable conditional phases allows us to engineer strong disorder in $\phi_i$ to fulfill this key requirement. 

We first measure the hallmark of an MBL-DTC: the persistent oscillation of local qubit polarizations $\braket{\hat{Z}(t)}$ at a period twice that of $\hat{U}_\text{F}$, irrespective of the initial state~\cite{Khemani2016, Else2016, Khemani2019, Ippoliti2020}. This subharmonic response is probed using a collection of random bit-string states, e.g. $\ket{01011...}$ where $0$($1$) denotes a single-qubit ground(excited) state in the $z$ basis. For each bit-string state, we generate a random instance of $\hat{U}_\text{F}$, and then measure $\braket{\hat{Z} (t)}$ every cycle. Figure~\ref{fig:2}b shows $\braket{\hat{Z} (t)}$ in a few different instances for a qubit near the center of the chain, $Q_{11}$, measured with $g = 0.60$ and $g = 0.97$. The former is deep in the thermal phase and indeed we observe rapid decay of $\braket{\hat{Z} (t)}$ toward 0 within 10 cycles for each instance. In contrast, for $g = 0.97$, $\braket{\hat{Z} (t)}$ shows large period-doubled oscillations persisting to over 100 cycles, suggestive of an MBL-DTC phase. The disorder averaged autocorrelator, $\avA = \overline{\braket{\hat{Z} (0) \hat{Z} (t)}}$, shows similar features (Fig.~\ref{fig:2}c).

We note that the data for $g=0.97$ is modulated by a gradually decaying envelope, which may arise from either external decoherence or slow internal thermalization\cite{LukinExp, Ho2017}. To establish DTC, additional measurements are needed to distinguish between these two mechanisms. This is achieved via an ``echo'' circuit $\hat{U}_\text{ECHO} = (\hat{U}_F^\dagger)^t \hat{U}_F^t$ which reverses the time evolution after $t$ steps. Deviations of $\hat{U}_\text{ECHO}$ from the identity operation are purely due to decoherence, and can be quantified via decay of the autocorrelator $A_0 \equiv (\langle \hat{Z} \hat{U}_\text{ECHO}^\dagger \hat{Z} \hat{U}_\text{ECHO} \rangle )^{1/2}$ (the square root accounts for the fact that $\hat{U}_\text{ECHO}$ acts twice as long as $\hat{U}_F^t$). A similar time-reversal technique was recently used in the study of out-of-time-ordered commutators in thermalizing random circuits~\cite{Mi_OTOC_2021}. 

Comparison between the disorder averaged $\avnorm$ and $\avA$ reveals qualitatively different behaviors in the two phases (Fig.~\ref{fig:2}c). In the thermal phase $g = 0.60$, $\avA$ approaches 0 much more quickly than $\avnorm$ does, indicating that the observed decay of $\avA$ is mostly induced by intrinsic thermalization. In the MBL-DTC phase $g = 0.97$, $\avnorm$ nearly coincides with the envelope of $\avA$, suggesting that decay of the latter is primarily induced by decoherence. The reference signal $\avnorm$ may be used to normalize $\avA$ and reveal its ideal behavior: $\avA / \avnorm$, shown in the upper panels of Fig.~\ref{fig:2}d, decays rapidly for $g = 0.60$ but retains near-maximal amplitudes for $g = 0.97$. Similar contrast between the two phases is seen in the error-mitigated autocorrelators $\avA / \avnorm$ for all qubits (bottom panel of Fig.~\ref{fig:2}d). The observation of a stable noise-corrected sub-harmonic response is suggestive of an MBL-DTC phase. 

\begin{figure}[t!]
	\centering
	\includegraphics[width=1\columnwidth]{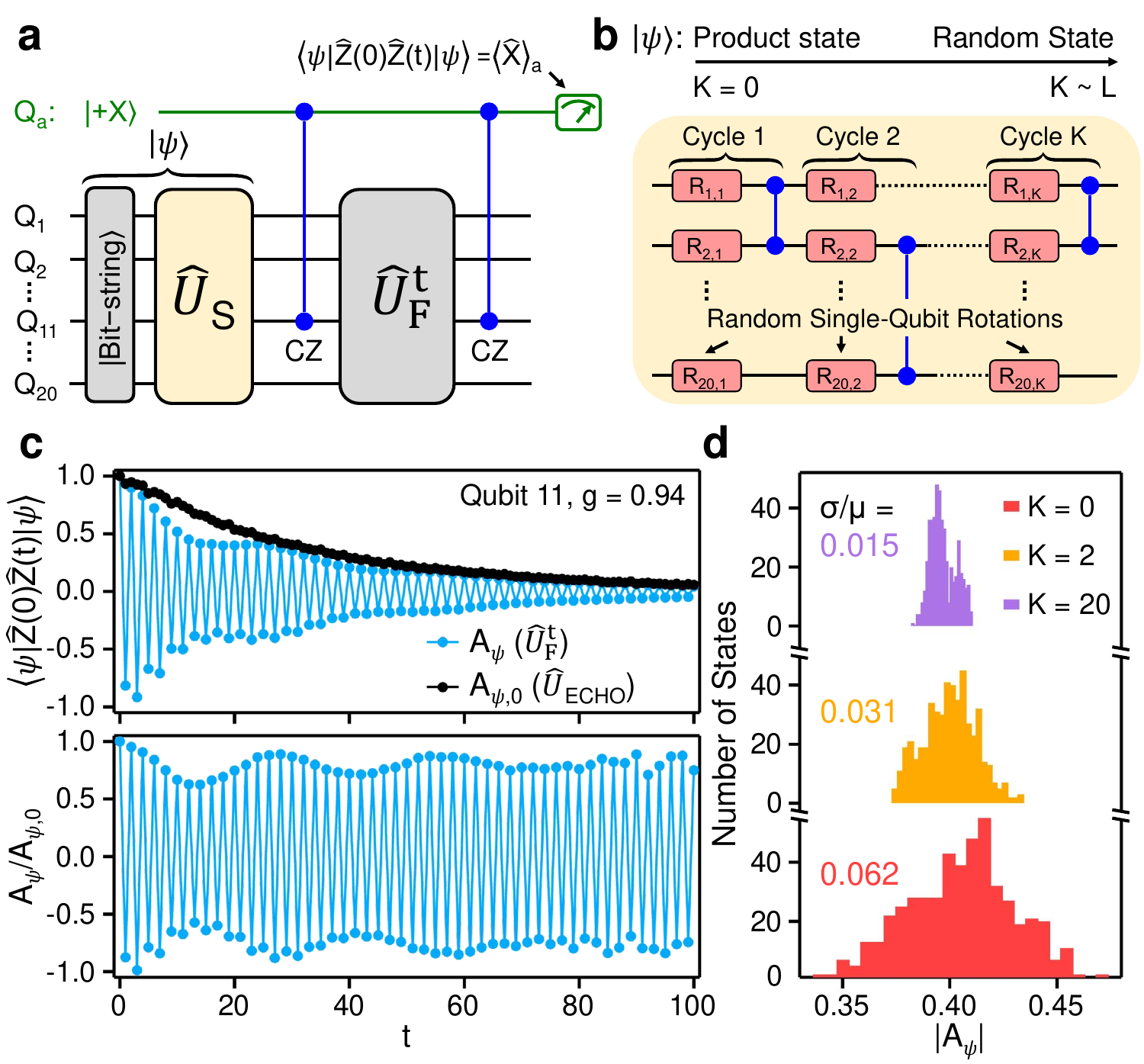}
	\caption{\textbf{Probing average spectral response via quantum typicality.} \textbf{a,} Schematic for measuring the auto-correlator, $A_\psi = \bra{\psi} \hat{Z} (0) \hat{Z} (t) \ket{\psi}$, on $Q_{11}$, of a scrambled quantum state $\ket{\psi}$. $\ket{\psi}$ is created by passing a bit-string state through a scrambling circuit, $\hat{U}_\text{S}$. An ancilla qubit $Q_\text{a}$ prepared in $\ket{+X}$ interacts with one of the qubits ($Q_{11}$) via a CZ gate before and after $\hat{U}_\text{F}^t$. The $x$-axis projection of $Q_\text{a}$, $\braket{\hat{X}}_\text{a}$, is measured at the end. \textbf{b,}  $\hat{U}_\text{S}$ contains $K$ layers of CZ gates interleaved with random single-qubit rotations, $R_{i, k}$, around a random axis along the equatorial plane of the Bloch sphere by an angle $\in [0.4 \pi, 0.6 \pi]$. \textbf{c,} Upper panel: $A_\psi$ for a single disorder instance with $K = 20$ cycles in $\hat{U}_\text{S}$. The square-root of the autocorrelator obtained by replacing $\hat{U}_\text{F}^t$ with $\hat{U}_\text{ECHO}$, $A_{\psi, 0}$, is also shown. Bottom panel: Normalized autocorrelator, $A_\psi / A_{\psi, 0}$, as a function of $t$. \textbf{d,} Histograms of $|A_{\psi}|$ from a single disorder instance, averaged over cycles 30 and 31. Each histogram corresponds to a different number of scrambling cycles, $K$, and includes data from 500 random initial bit-string states fed through the scrambling circuit.}
	\label{fig:4}
\end{figure}

We now turn to a systematic analysis of the next requirement necessary to establish an MBL-DTC: namely the presence of eigenstate order across the entire spectrum which, in turn, implies that sub-harmonic response should not be strongly affected by the choice of initial states. 
In contrast, various prethermal mechanisms in driven systems predict strong dependence of the thermalization rate on the initial state, e.g. through its quantum numbers~\cite{BarrettExp, Luitz2020} or its energy under an effective time-independent Hamiltonian $\hat{H}_\text{eff}$~\cite{Else2017, Abanin2017, Mori_2018} that approximately governs the dynamics for small system sizes and/or finite times.
To elucidate this aspect of the MBL-DTC phase, we analyze in detail the distribution of autocorrelator values over initial bit-string states. 

We begin by examining the position- and disorder-averaged autocorrelator $\avavA$ over three representative bit-string initial states, shown in the left panel of Fig~\ref{fig:3}a. The square brackets indicate averaging over qubits in the chain (excluding the edge qubits $Q_1$, $Q_{20}$, which may be affected by the presence of edge modes independent of the bulk DTC response~\cite{MitraEdge}). The three time traces are nearly indistinguishable. This behavior is in clear contrast with a model without eigenstate order, implemented by a family of drives $\hat{U}'_\text{F}$ where the $\phi_i$ angles are set to a uniform value~\footnote{This value of $\phi_i=-0.4$ is chosen to be small enough that a leading order high-frequency Floquet Magnus expansion to obtain $\hat{H}_\text{eff}$ is a reasonable approximation (see SM).}, $\phi_i = -0.4$.
Without disorder in the $\phi_i$, the drive $\hat{U}'_\text{F}$ is not asymptotically localized but exhibits transient DTC-like behavior. 
Here, $\avavA$ for $\hat{U}'_\text{F}$ (disorder averaged over random $h_i$ only), shown in the right panel of Fig~\ref{fig:3}a, reveals markedly different decay rates for the three states. The random bit-string state, in particular, decays faster than the polarized or N\'eel states. 

A more comprehensive analysis, presented in Fig.~\ref{fig:3}b, is based on sampling the absolute values of $\avavA$ for 500 random initial bit-string states (averaged over cycles 30 and 31).
For the MBL-DTC $\hat{U}_F$, the histogram is tight, with a relative standard deviation (ratio of standard deviation to mean, $\sigma / \mu$) of 0.038. Here the non-zero value of $\sigma$ likely stems from finite experimental accuracy and number of disorder instances, as analysis in the SM shows that $\avavA$ is independent of the initial state. In contrast, the $\hat{U}_F'$ model shows a broader distribution with a much lower mean, and with $\sigma / \mu = 0.129$. Moreover, the histogram is asymmetrical, with outliers at high $\avavA$ including the polarized and N\'eel states (51\% and 88\% higher than the mean, respectively). 
These two states are special because they are low temperature states that sit near the edge of the spectrum of   $\hat{H}_\text{eff}$ (see SM).
Plotting the autocorrelator $\avavA$ against the energy of each bitstring under $\hat{H}_\text{eff}$, in the inset of Fig.~\ref{fig:3}b, reveals a clear correlation.
No such correlation is present in the MBL model.

Independent confirmation of MBL as the mechanism underlying the stability of DTC is achieved by characterizing the propagation of correlations. In MBL dynamics, local perturbations spread only logarithmically in time~\cite{AbaninRMP2019}, as opposed to algebraic ($\sim t^\alpha$) spreading in thermalizing dynamics. We prepare two initial bitstring states differing by only a single bit-flip at $Q_{11}$ and measure  $\braket{\hat{Z} (t)}$ for each site in both states (Fig.~\ref{fig:3}c). It can be seen that the difference in the two signals, $\zeta_1$ and $\zeta_2$, decays rapidly with the distance from $Q_{11}$ for disordered $\phi_i$ and becomes undetectable at $Q_{14}$. On the other hand, for uniform $\phi_i = -0.4$, $\zeta_1$ and $\zeta_2$ have a much more pronounced difference which remains significant at $Q_{14}$. This difference is further elucidated by the ratio $\zeta_\text{r} = |\zeta_1-\zeta_2|/(|\zeta_1|+|\zeta_2|)$, shown in Fig.~\ref{fig:3}d. Physically, $\zeta_\text{r}$ corresponds to the relative change in local polarization as a result of the bit flip, and is inherently robust against qubit decoherence (see SM). We observe that up to $t = 100$, $\zeta_\text{r}$ remains sharply peaked around the initial perturbation ($Q_{11}$) for disordered $\phi_i$. In contrast, a propagating light cone is visible for $\phi_i = -0.4$, with the perturbation reaching all qubits across the chain as $t$ increases. The spatial profiles of $\zeta_\text{r}$ at $t = 51$ to $t = 60$ (right panel of Fig.~\ref{fig:3}d) show that $\zeta_\text{r}$ is much sharper for disordered $\phi_i$. 
This slow propagation provides strong indication of MBL and another experimental means of distinguishing eigenstate-ordered phases from transient phenomena. 

Our measurement of $\avavA$ for 500 initial states in Fig.~\ref{fig:3}d provides clear evidence of initial state independence. Still, a direct sampling of states is practically limited to small fractions of the computational basis (0.05\% in this case) and would suffer from the exponential growth of the Hilbert space on larger systems. A more scalable alternative is to use random, highly entangled states to directly measure spectrally-averaged quantities (quantum typicality\cite{Popescu2006, Goldstein2006, Pal2021}). The autocorrelator $A$ averaged over all $2^L$ bitstrings agrees, up to an error exponentially small in $L$, with $A_\psi = \bra{\psi} \hat{Z}(0)\hat{Z}(t) \ket{\psi}$, where $\ket{\psi}$ is a typical Haar-random many-body state in the Hilbert space of $L$ qubits. We prepare such a state by evolving a bitstring with a random circuit $\hat{U}_S$ of variable depth $K$ (Fig.~\ref{fig:4}b), and couple an ancilla qubit to the system to measure the two-time operator $\hat{Z}(0)\hat{Z}(t)$ (Fig.~\ref{fig:4}a). Experimental results for the error-mitigated, spectrally averaged signal $A_\psi/A_{\psi,0}$ on qubit $Q_{11}$ (Fig.~\ref{fig:4}c) show behavior consistent with a stable MBL-DTC. The effect of the state-preparation circuit $\hat{U}_S$ is illustrated by the dependence of the relative standard deviation $\sigma / \mu$ for $A_\psi$ on $K$. As shown in Fig.~\ref{fig:4}d, $\sigma / \mu$ steadily decreases as $K$ increases, reducing from a value of 0.062 at $K = 0$ to a value of $0.015$ at $K = 20$. This is consistent with the fact that  $|\psi\rangle$ becomes closer to a Haar-random state as $K$ increases. We use a single disorder instance to study the convergence of the quantum typicality protocol because disorder averaging independently leads to narrow distributions even for $K=0$ (Fig.~\ref{fig:3}b).

The scaling with $L$ of the spectrally-averaged autocorrelator, at a time $t\sim {\rm poly}(L)$, provides a sharp diagnostic: this saturates to a finite value in the MBL-DTC, while it scales to zero with increasing $L$ in transient cases (where, for instance, a vanishing fraction of the spectrum of an appropriate $\hat{H}_\text{eff}$ shows order). While the averaged autocorrelator may be unduly affected by outlier states and/or long (but $O(1)$) thermalization times at small system sizes and times (thereby making the complementary bitstring analysis of Fig.~\ref{fig:3} essential), the polynomial scaling of this protocol establishes a proof of principle for efficiently verifying the presence or absence of an MBL DTC in a range of models as quantum processors scale up in size to surpass the limits of classical simulation~\cite{Arute2019}.

\begin{figure}[t!]
	\centering
	\includegraphics[width=1\columnwidth]{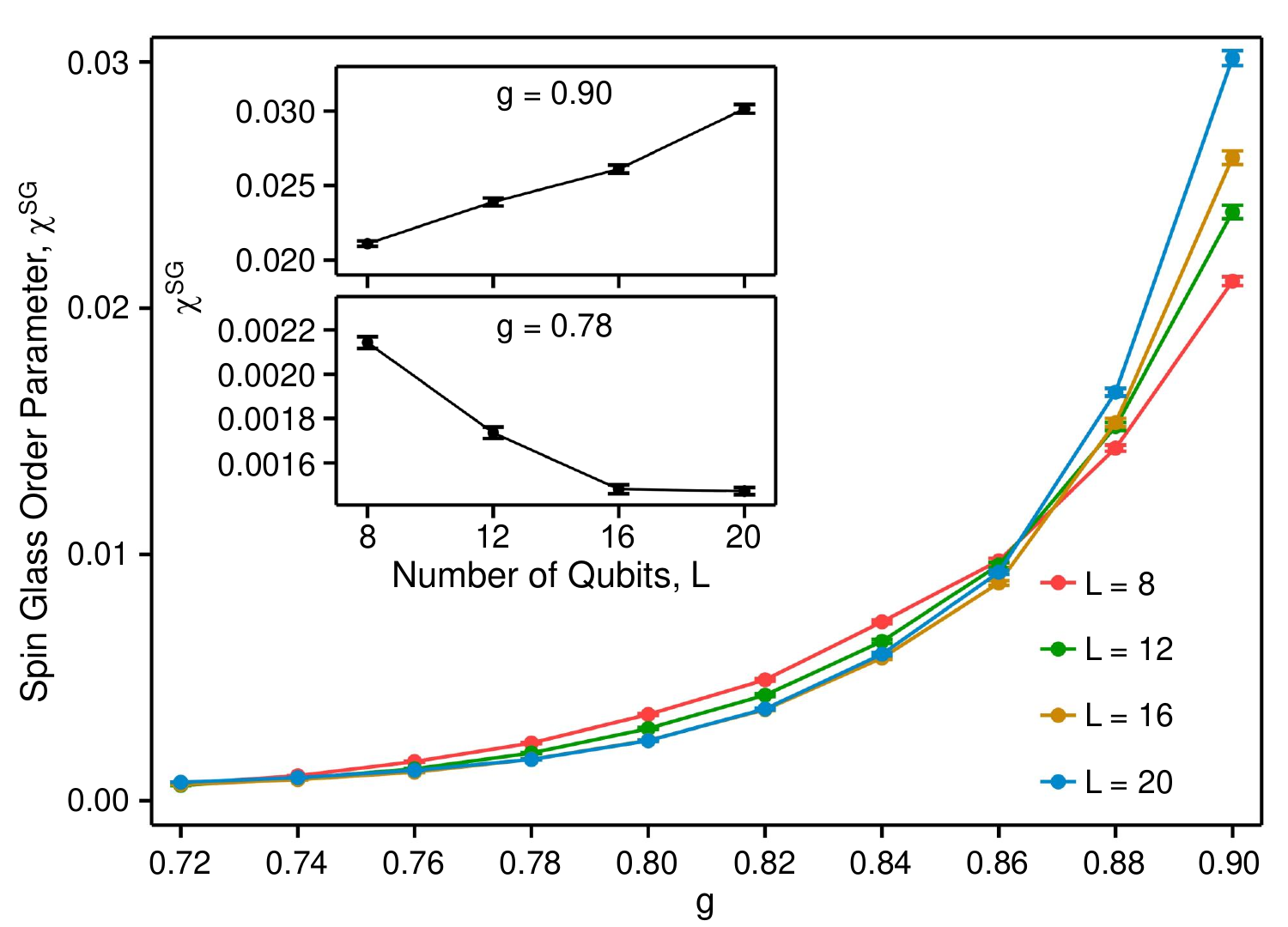}
	\caption{\textbf{Estimating phase-transition by varying system size.} Spin-glass order parameter $\chi^\text{SG}$ as a function of $g$ for different chain lengths $L$, measured between $t=51$ and $t=60$.  Every data point is averaged over 40 disorder instances. To construct $\chi^\text{SG}$, we sample 40000 bit-strings at the output of $\hat{U}_\text{F}^t$ for each cycle and disorder instance. To address the inhomogeneity of qubit coherence, smaller qubit chains are also averaged over different possible combinations of qubits. For example, $L = 12$ is averaged over 12-qubit chains made from $Q_1$ through $Q_{12}$, $Q_3$ through $Q_{15}$ etc. Error bars correspond to statistical errors (estimated by resampling data from the 40 disorder instances via the jackknife method). Contribution of hardware (e.g. gate) errors to $\chi^\text{SG}$ is not included in the error bars. Inset shows the size-dependence of $\chi^\text{SG}$ for two different values of $g$.}
	\label{fig:5}
\end{figure}

Finally, we systematically vary $g$ in small increments and obtain an experimental finite-size analysis to establish the extent of the MBL phase and the transition out of it. Sharply defining phases of matter, whether in or out of equilibrium, requires a limit of large system size. 
Thus it is important to examine the stability of the MBL-DTC and thermalizing regimes observed in our finite-size quantum processor as the size of the system is increased. 
To address this, we measure an Edwards-Anderson spin glass order parameter\cite{Edwards1975, Kjall2014}, 
\begin{equation}
\chi^{\text{SG}} = \frac{1}{L-2} \sum_{i\neq j}{}^{'} \braket{\hat{Z}_i\hat{Z}_j}^2
\end{equation}
(the primed sum excludes edge qubits $Q_1$, $Q_L$), as a function of time. This quantity measures the persistence of random (``glassy'') spatial patterns in the initial bitstring state: at late times, it vanishes with increasing $L$ in the thermalizing phase $g<g_c$, while it is extensive in the MBL-DTC $g>g_c$.
As a result, it is expected to show a finite-size crossing at $g\simeq g_c$ (though the precise location is subject to strong finite-size and finite-time drifts\cite{Pal2010, Abanin2019}).
Experimentally, $\chi^{\text{SG}}$ is constructed from bit-string samples obtained by jointly reading out all qubits and then averaged over cycles and disorder instances (Fig.~\ref{fig:5}). The size of the qubit chain is varied by restricting the drive $\hat{U}_F$ to contiguous subsets of 8, 12, and 16 qubits (as well as the entire 20-qubit chain). 
We observe increasing (decreasing) trends in $\chi^{\text{SG}}$ vs $L$ when $g$ is above (below) a critical value $g_c$.
The data indicate $0.83 \lesssim g_c\lesssim 0.88$, consistent with numerical simulations (see SM).

In conclusion, we have demonstrated the possibility of engineering and characterizing non-equilibrium phases of matter on a quantum processor, providing direct experimental observation of an eigenstate-ordered MBL-DTC. 
The scalability of our protocols sets a blueprint for future studies of non-equilibrium phases and phase transitions on complex quantum systems beyond classical simulability. 
The efficient {verification} of eigenstate order 
can inspire a general strategy for establishing whether a desired property, such as a particular phase, is in fact present in a quantum processor. 

\onecolumngrid

\vspace{1em}
\begin{flushleft}
{\hypertarget{authorlist}{${}^\dagger$} \small Google Quantum AI and Collaborators}

\bigskip
{\small

\renewcommand{\author}[2]{#1$^\textrm{\scriptsize #2}$}
\newcommand{\firstauthor}[2]{#1$^{\textrm{\scriptsize #2}, \hyperlink{equal}{*}}$}
\newcommand{\corrauthora}[2]{#1$^{\textrm{\scriptsize #2}, \hyperlink{corra}{\ddagger}}$}
\newcommand{\corrauthorb}[2]{#1$^{\textrm{\scriptsize #2}, \hyperlink{corrb}{\mathsection}}$}
\renewcommand{\affiliation}[2]{$^\textrm{\scriptsize #1}$ #2 \\}

\newcommand{\xGoogle}{\affiliation{1}{Google Research, Mountain View, CA, USA}}

\newcommand{\xStanford}{\affiliation{2}{Department of Physics, Stanford University, Stanford, CA, USA}}

\newcommand{\xMIT}{\affiliation{3}{Department of Physics, Massachusetts Institute of Technology, Cambridge, MA, USA}}

\newcommand{\xUMass}{\affiliation{4}{Department of Electrical and Computer Engineering, University of Massachusetts, Amherst, MA, USA}}

\newcommand{\xUCSB}{\affiliation{5}{Department of Physics, University of California, Santa Barbara, CA, USA}}

\newcommand{\xUCR}{\affiliation{6}{Department of Electrical and Computer Engineering, University of California, Riverside, CA, USA}}

\newcommand{\xPritzker}{\affiliation{7}{Pritzker School of Molecular Engineering, University of Chicago, Chicago, IL, USA}}
\newcommand{\xColumbia}{\affiliation{8}{Department of Chemistry, Columbia University, New York, NY 10027, USA}}

\newcommand{\xPrinceton}{\affiliation{9}{Department of Physics, Princeton University, Princeton, NJ, USA}}

\newcommand{\xMaxPlanck}{\affiliation{10}{Max-Planck-Institut f\"{u}r Physik komplexer Systeme, 01187 Dresden, Germany}}

\newcommand{\Google}{1}
\newcommand{\stanford}{2}
\newcommand{\MIT}{3}
\newcommand{\UMass}{4}
\newcommand{\UCSB}{5}
\newcommand{\UCR}{6}
\newcommand{\Pritzker}{7}
\newcommand{\Columbia}{8}
\newcommand{\Princeton}{9}
\newcommand{\MaxPlanck}{10}

\firstauthor{Xiao Mi}{\Google},
\firstauthor{Matteo Ippoliti}{\stanford},
\author{Chris Quintana}{\Google},
\author{Ami Greene}{\Google,\! \MIT},
\author{Zijun Chen}{\Google},
\author{Jonathan Gross}{\Google},
\author{Frank Arute}{\Google},
\author{Kunal Arya}{\Google},
\author{Juan Atalaya}{\Google},
\author{Ryan Babbush}{\Google},
\author{Joseph C.~Bardin}{\Google,\! \UMass},
\author{Joao Basso}{\Google},
% ...

\vspace{4mm}
\xGoogle
\xStanford
\xMIT
\xUMass
\xUCSB
\xUCR
\xPritzker
\xColumbia
\xPrinceton
\xMaxPlanck
}

\vspace{3mm}
\begin{footnotesize}
{\hypertarget{equal}{${}^*$} These authors contributed equally to this work.}\\
\end{footnotesize}

\vspace{3mm}
\begin{footnotesize}
{\bf Acknowledgements---}This work was supported in part from the Defense Advanced Research Projects Agency (DARPA) via the DRINQS program (M.I., V.K., R.M., S.S.), by a Google Research Award: Quantum Hardware For Scientific Research In Physics (V.K. and M.I.), and from the Sloan Foundation through a Sloan Research Fellowship (V.K.) The views, opinions and/or findings expressed are those of the authors and should not be interpreted as representing the official views or policies of the Department of Defense or the U.S. Government.
M.I. was funded in part by the Gordon and Betty Moore Foundation's EPiQS Initiative through Grant GBMF8686. 
This work was partly supported by the Deutsche Forschungsgemeinschaft under grants SFB 1143 (project-id 247310070) and the cluster of excellence ct.qmat (EXC 2147, project-id 390858490).\par 
\end{footnotesize}

\vspace{3mm}
\begin{footnotesize}
{\bf Author contributions---}M.I., K.K., V.K., R.M. and S.S. conceived the project. X.M., C.Q. and P.R. executed the experiment. All the aforementioned discussed the project in progress and interpreted the results.  M.I., K.K., V.K. and X.M., designed measurement protocols. J.C., A.G., J.G. and X.M. implemented and calibrated the CPHASE gates. M.I. and V.K. performed theoretical and numerical analyses.  M.I., K.K., V.K., R.M., X.M., and P.R. wrote the manuscript.  M.I. and X.M. wrote the supplementary material.  Y.C.,  K.K., V.K., H.N., P.R. and V.S. led and coordinated the project. Infrastructure support was provided by Google Quantum AI. All authors contributed to revising the manuscript and the supplementary material.\par 
\end{footnotesize}

\vspace{3mm}
\begin{footnotesize}
{\bf Data availability---}The data presented in this work are available from the corresponding authors upon reasonable request.\par 
\end{footnotesize}

\end{flushleft}

\twocolumngrid
\bibliography{DTC_Refs}

\clearpage


\newpage

\onecolumngrid
\begin{center}
    \textbf{\large Supplementary Materials for ``Observation of Time-Crystalline\\ Eigenstate Order on a Quantum Processor''}
\end{center}
\vspace{18pt}
\twocolumngrid

\section{CPHASE Gate Implementation and Error Benchmarking}

An essential building block for the quantum circuits used to observe many-body localized DTC is the two-qubit gate $ZZ (\phi) = e^{-i\frac{\phi}{4} \hat{Z}_\text{a} \hat{Z}_\text{b}}$, which belongs to the more general family of Fermionic Simulation (FSIM) gates having the unitary form $\hat{U}_\text{FSIM}$ \cite{Arute2019}:

\begin{equation}
\begin{pmatrix}
1 & 0 & 0 & 0 \\
0 & e^{i (\Delta_+ + \Delta_-)} \cos{\theta} & -ie^{i (\Delta_+ - \Delta_{-, \text{off}})} \sin{\theta} & 0 \\
0 & -ie^{i (\Delta_+ + \Delta_{-, \text{off}})} \sin{\theta} & e^{i (\Delta_+ - \Delta_-)} \cos{\theta} & 0 \\
0 & 0 & 0 & e^{i (2 \Delta_+ - \phi)}
\end{pmatrix}.
\end{equation}
Here $\theta$ is the two-qubit iSWAP angle and $\Delta_+$, $\Delta_-$ and $\Delta_{-, \text{off}}$ are phases that can be freely adjusted by single-qubit $Z$-rotations. In this parametrized representation, $ZZ (\phi) = \hat{U}_\text{FSIM} (\theta = 0, \Delta_- = 0, \Delta_{-, \text{off}} = 0, \phi = 2 \Delta_+)$, which is equivalent to a CPHASE gate with conditional-phase $\phi$ and a single-qubit rotation $Z (\frac{\phi}{2})$ acting on each qubit. Precise single-qubit $Z$-control has already been demonstrated in our previous work \cite{Mi_OTOC_2021}. Here, we focus on implementing CPHASE gates with a variable $\phi$.

\begin{figure}[t!]
	\centering
	\includegraphics[width=1\columnwidth]{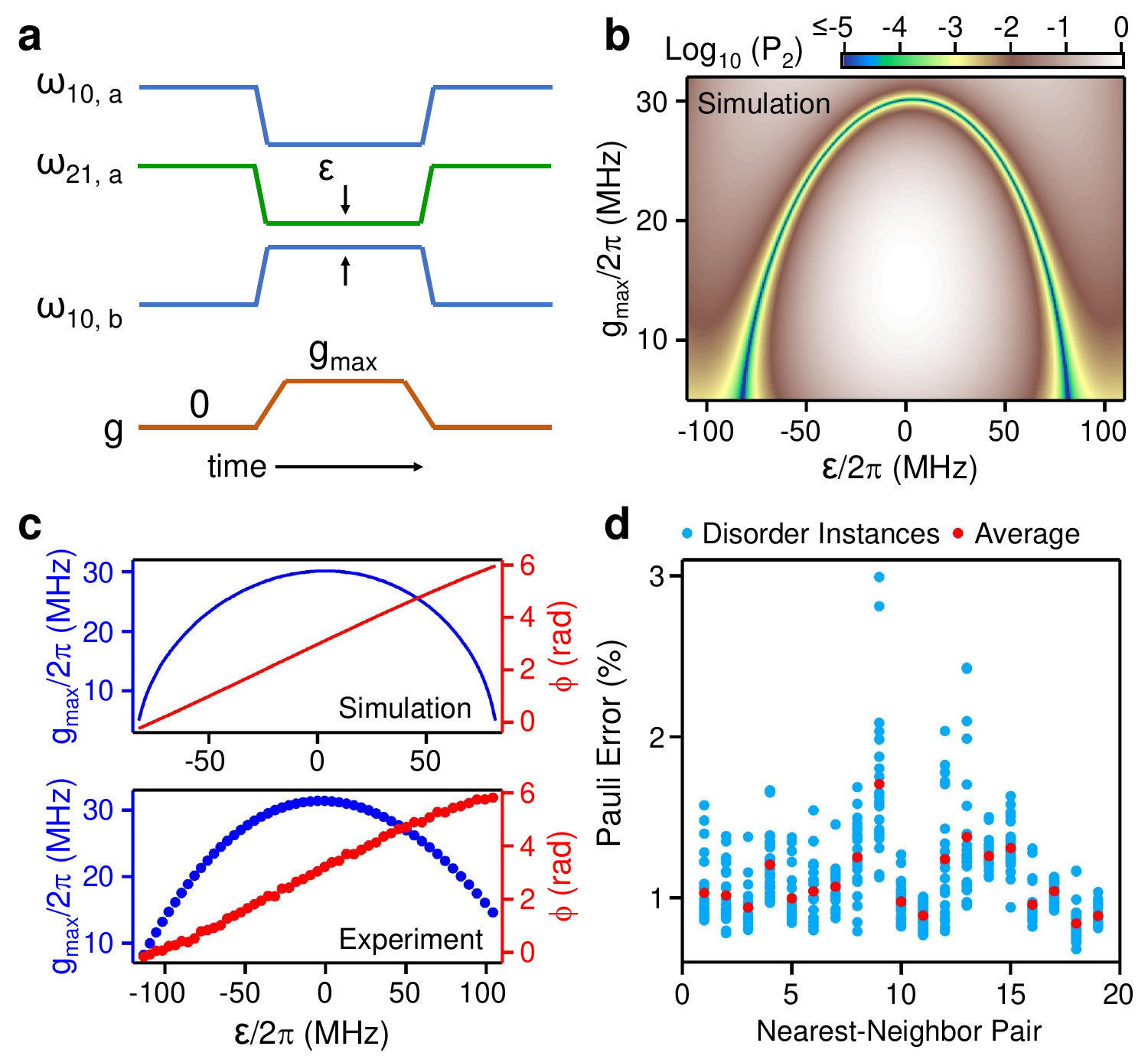}
	\caption{\textbf{Implementing CPHASE gates with tunable transmon qubits.} \textbf{a,} Schematic of flux pulses used to realize a CPHASE gate. The frequencies of two coupled transmons, $\omega_\text{10, a}$ and $\omega_\text{10, b}$, are displaced from their idle positions into a configuration wherein $\omega_\text{10, b}$ is detuned from $\omega_\text{21, a}$ by an amount $\epsilon$. At the same time, a flux pulse on the coupler turns on an inter-qubit coupling $g > 0$ with a maximum value of $g_\text{max}$ for a fixed duration of $\sim$18 ns. \textbf{b,} Simulated values of leakage, $P_2$, as a function of $\epsilon$ and $g_\text{max}$, using typical device parameters and pulse shapes. \textbf{c,} The values of $g_\text{max}$ and conditional-phase $\phi$ at the locations of minimum leakage, plotted for different values of $\epsilon$. Upper panel shows simulated results and lower panel shows representative experimental values obtained from one pair of qubits. \textbf{d,} Pauli error rates for each neighboring pair of qubits in the 20-qubit chain used by the experiment, obtained from parallel XEB. Each error rate includes contributions from two random single-qubit gates ($\pi/2$ rotations around a random axis along the equatorial plane of the Bloch sphere) and a CPHASE gate. Data are shown for 24 disorder instances, with each instance including a different random set of $\phi_i$ across the qubit chain. Red dots show the average error of each qubit pair.}
	\label{fig:1s}
\end{figure} 

The qubits used in our experiment are superconducting transmon qubits with both tunable frequencies and tunable inter-qubit couplings. Due to the existence of higher states, a natural way to realize a CPHASE gate is to bring the $\ket{11}$ and $\ket{02}$ states of two coupled qubits close to resonance diabatically, allow the qubits to interact for a duration $\sim\frac{1}{\sqrt{8 g^2 + \epsilon^2}}$, before turning off the inter-qubit coupling and ramping the qubits back to their idle configurations. Here $\ket{0}$ and $\ket{1}$ are the ground and excited states of each qubit, $\ket{2}$ is a leakage state outside the computational space, $g$ denotes the inter-qubit coupling (between the $\ket{10}$ and $\ket{01}$ states) and $\epsilon$ is the detuning between the $\ket{11}$ and $\ket{02}$ states during interaction. A schematic for the flux pulses to realize the CPHASE gate is shown in Fig.~\ref{fig:1s}a. 

Figure~\ref{fig:1s}b shows simulated values of leakage, namely the probability of one qubit residing in $\ket{2}$ after the CPHASE gate, as a function of $\epsilon$ and maximum value of $g$ during interaction, $g_\text{max}$. A narrow arc-like region, corresponding to a contour $8 g_\text{max}^2 + \epsilon^2 = $ constant, can be seen from the simulation. The values of $g_\text{max}$ and $\phi$ along this contour are shown in the upper panel of Fig.~\ref{fig:1s}c, where we see a full range of $\phi \in [-2 \pi, 0]$ can be achieved for absolute detuning values of $|\epsilon| / 2 \pi < 100$ MHz. Since the $\ket{01}$ and $\ket{10}$ states are detuned by $\sim$200 MHz, their interaction remains dispersive during the CPHASE gate and therefore ensures a small iSWAP angle $\theta$ (we confirm this experimentally in the next section).

Experimentally, the leakage-minimizing value of $g_\text{max}$ is detected for a discrete set of $\epsilon$ via $\ket{2}$ state readout and the corresponding $\phi$ is then calibrated using unitary tomography \cite{Foxen_PRL_2020}. A polynomial fit is then performed to infer values of $\epsilon$ and $g_\text{max}$ for intermediate values of $\phi$, thereby achieving a continuous family of CPHASE gates with fully tunable $\phi$. Example experimental calibration data for $\epsilon$, $g_\text{max}$ and $\phi$ are included in the bottom panel of Fig.~\ref{fig:1s}c. Excellent agreement with numerical results is found. The discrepancy in values of $\epsilon$ likely arises from deviation between the actual pulse shapes on the quantum processor and those used in the numerical simulation.

To estimate the errors of typical CPHASE gates, we perform cross-entropy benchmarking (XEB) similar to previous works \cite{Arute2019, Mi_OTOC_2021}. Here the gates are characterized in parallel and therefore include errors arising from any cross-talk effects. The XEB results for 24 random combinations of $\phi_i$ across the 20-qubit chain used in the main text are shown in Fig~\ref{fig:1s}d, where we have used the so-called ``cycle'' Pauli error as the metric. A cycle Pauli error includes errors from two single-qubit gates and a single CPHASE gate. In computing the XEB fidelities, we have also assumed the CPHASE gate with the calibrated $\phi$ as the ideal gate \cite{Mi_OTOC_2021}. As such, the Pauli errors include contributions from both incoherent effects such as qubit relaxation and dephasing, as well as coherent effects such as any mis-calibration in $\phi$ and residual values of $\theta$. The error rates are observed to be relatively dependent on $\phi$, a likely consequence of changes in coherent and incoherent errors when the pulse parameters are varied. Overall, we observe an average error rate of 0.011, comparable to gates used in our past works \cite{Arute2019, Mi_OTOC_2021}.

\section{Floquet Calibration of CPHASE Gates}

After basic pulse-level calibration of CPHASE gates, the $ZZ (\phi)$ gate is then calibrated using the technique of Floquet calibration \cite{zhang_floquet_2020,Neill_Nature_2021}. Floquet calibration utilizes periodic circuits which selectively amplify different parameters within $\hat{U}_\text{FSIM}$, allowing for sensitive detection and rectification of small coherent errors for such quantum gates. Our past works have primarily focused on calibrating the iSWAP-like family of gates, where $\theta \gg \phi$. For $ZZ$ gates, the opposite limit $\phi \gg \theta$ is true and the optimal calibration circuits are in some cases different from our previous works. In this section, we present the gate sequences and example calibration results for the $ZZ$ gates. For a detailed description of the underlying theory of Floquet calibration, the reader is directed to our previous publications \cite{zhang_floquet_2020,Neill_Nature_2021}.

\subsection{Calibration of $\Delta_+$, $\Delta_-$ and $\phi$}

\begin{figure}[t!]
	\centering
	\includegraphics[width=1\columnwidth]{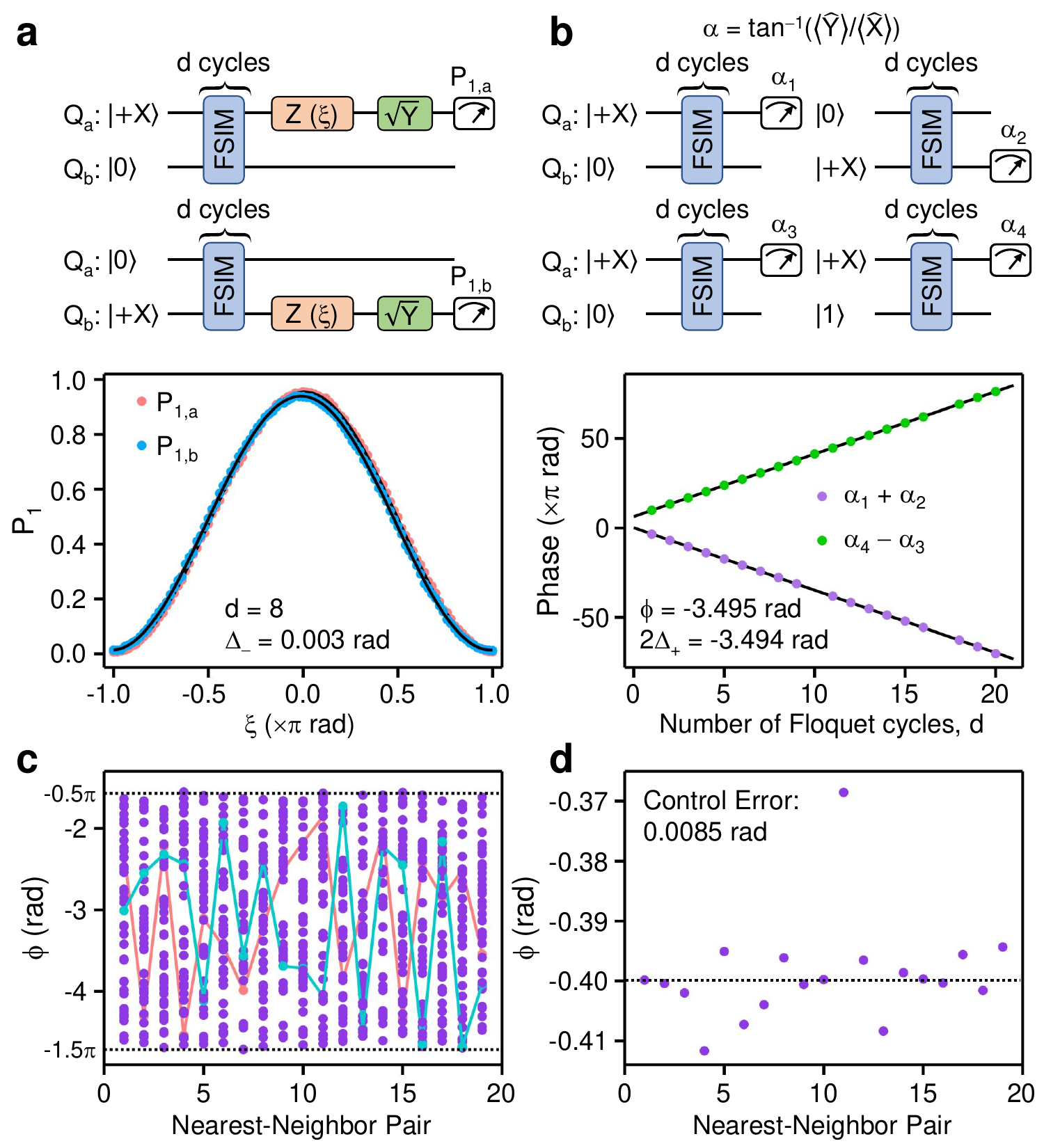}
	\caption{\textbf{Floquet calibration of single-qubit and conditional phases for CPHASE-like gates.} \textbf{a,} Top panel: Gate sequences for calibrating the $\Delta_-$ angle of the FSIM. Bottom panel: Example experimental data for $P_\text{1,a}$ and $P_\text{1,b}$ ($\ket{1}$ state population for qubits $Q_\text{a}$ and $Q_\text{b}$, respectively) as functions of $\xi$. The circuit depth is fixed at $d = 8$. Solid black lines show fits to a cosine function for each qubit, which allow $\Delta_-$ to be extracted. \textbf{b,} Top panel: Gate sequences for calibrating $\Delta_+$ and $\phi$. For each of the 4 gate sequences, the $\braket{\hat{X}}$ and $\braket{\hat{Y}}$ projections of the Bloch vector for a given qubit are measured at the end, from which an angle $\alpha = \tan^{-1} \left( \braket{\hat{Y}} / \braket{\hat{X}} \right)$ is computed. Bottom panel: Example experimentally obtained phase sums ($\alpha_1$ + $\alpha_2$) and differences ($\alpha_4$ - $\alpha_3$) as functions of $d$, number of cycles in the Floquet gate sequences. Solid black lines show linear fits, the slopes of which determine $\phi$ and $\Delta_+$. \textbf{c,} Experimentally measured $\phi$ for each neighboring pair of qubits in the 20-qubit chain. Results from 40 disorder instances are plotted. The blue and red connected dots indicate the values of two particular instances, while all other instances are shown as disconnected purple dots. Dashed lines corresponding to $\phi = -0.5 \pi$ and $\phi = -1.5 \pi$. \textbf{d,} Experimental measurements of $\phi$ when a target value is set at $-0.4$ (dashed line) for all nearest-neighbor pairs. An average deviation of 0.0085 rad is found between the target and measured values of $\phi$.}
	\label{fig:2s}
\end{figure} 

\begin{figure}[t!]
	\centering
	\includegraphics[width=1\columnwidth]{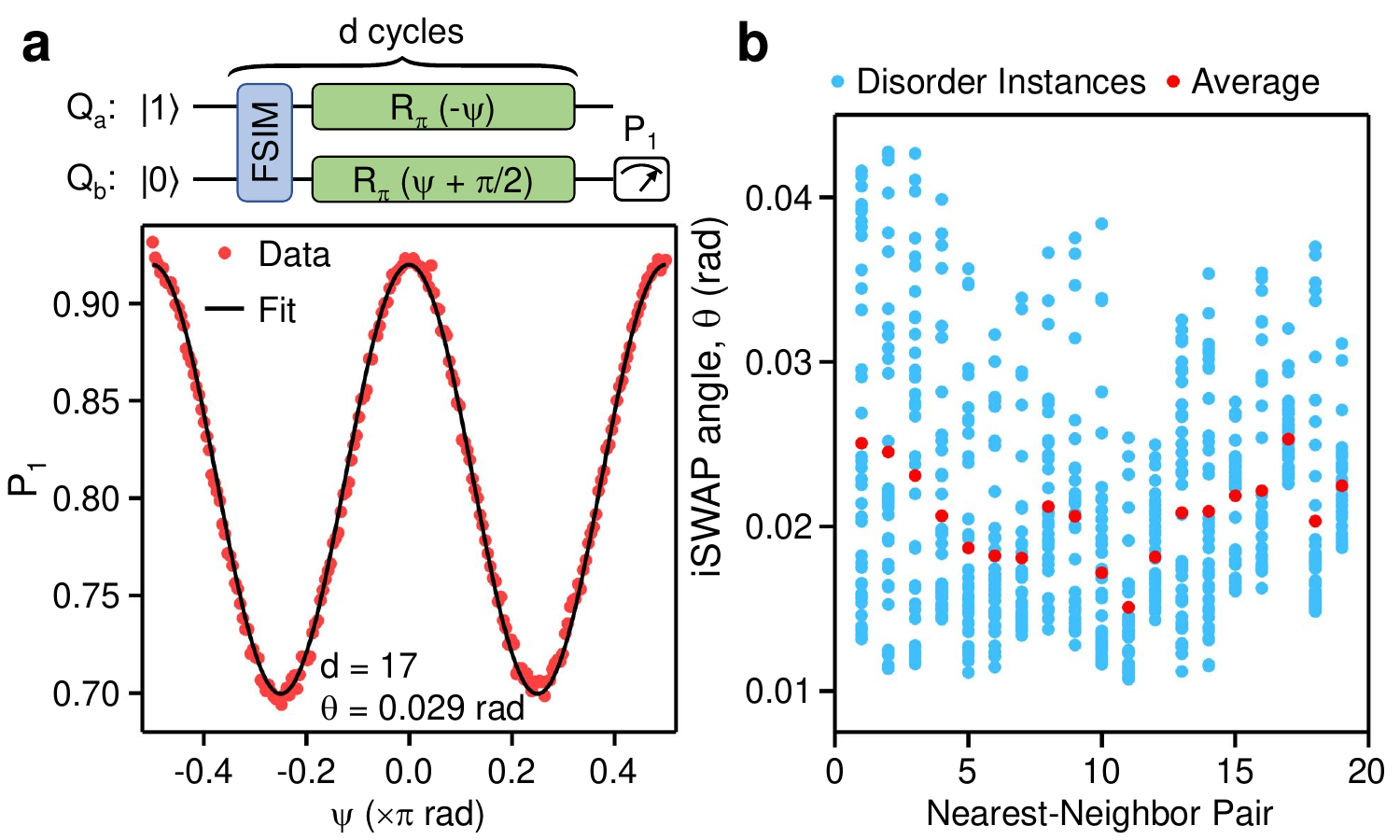}
	\caption{\textbf{Floquet calibration of small iSWAP angles.} \textbf{a,} Top: Periodic circuit for calibrating $\theta$: Each cycle includes an FSIM gate, followed by two single-qubit rotations $R_\pi (-\psi)$ and $R_\pi (\psi + \pi / 2)$. After $d$ cycles, the excited state population $P_1$ of $Q_\text{b}$ is measured. Bottom: Example experimental data at a fixed depth $d = 17$, showing $P_1$ as a function of $\psi$. Solid black line shows fit to a sinusoidal function, the amplitude of which determines the value of $\theta$. \textbf{b,} Experimentally measured $\theta$ for each neighboring pair of qubits in the 20-qubit chain. Results from 40 disorder instances are plotted as blue dots, and the average value for each qubit pair is plotted as red dots. Overall, $\theta$ has a mean value of 0.022 rad and a standard deviation 0.014 across all qubit pairs and disorder instances.} 
	\label{fig:3s}
\end{figure} 

The calibration circuits for $\Delta_-$ are illustrated in Fig.~\ref{fig:2s}a and comprise two Ramsey-like measurements: Each qubit is separately initialized along the x-axis of the Bloch sphere, $\ket{X}$. A total number of $d$ FSIM gates are then applied, which in general rotate the excited qubit around the z-axis of the Bloch sphere due to non-zero single-qubit phases within the uncalibrated $\hat{U}_\text{FSIM}$. At the end of the sequence, a $Z$ gate with a rotation angle $\xi$ is applied to the excited qubit, followed by a $\sqrt{Y}$ gate. The resulting $\ket{1}$ state population, $P_1$, is then measured. Example data for $P_1$ of each qubit are shown in the bottom panel of Fig.~\ref{fig:2s}a, which are fitted to a cosine function $P_1 = B_0 + B_1 \cos (\xi + \xi_0)$ where $B_0$, $B_1$ and $\xi_0$ are fitting parameters. The value of $\Delta_-$ is then equal to $\frac{\xi_a - \xi_b}{2d}$, where $\xi_a$ ($\xi_b$) is the fitted $\xi_0$ for $Q_\text{a}$ ($Q_\text{b}$). This equivalence may be understood through the fact that $2\Delta_-$ is the difference in the degree of local $Z$ rotations undergone by each qubit after the application of one FSIM gate between them.

The phases $\Delta_+$ and $\phi$ are calibrated using four periodic circuits sharing a similar structure, as indicated in the top panel of Fig.~\ref{fig:2s}b. For $\Delta_+$, we again separately prepare each qubit in the $\ket{X}$ state while leaving the other qubit in $\ket{0}$. The FSIM gate is then applied $d$ times. At the end of the sequence, two tomographic measurements are performed on the excited qubit to determine the angle of its Bloch vector, $\alpha = \tan^{-1} \left( \braket{\hat{Y}} / \braket{\hat{X}} \right)$. The total accumulated phase $\alpha_1 + \alpha_2$, where $\alpha_1$ ($\alpha_2$) is $\alpha$ for $Q_\text{a}$ ($Q_\text{b}$), is equivalent to $2d \Delta_+$. This equivalence arises from the fact that $2\Delta_+$ physically corresponds to the sum of the degrees of local $Z$ rotations on both qubits after the application of one FSIM gate.

The conditional-phase $\phi$ is calibrated by preparing one qubit ($Q_\text{a}$) in $\ket{X}$ and comparing $\alpha$ when the other qubit ($Q_\text{b}$) is initialized in either the $\ket{0}$ or the $\ket{1}$ state. A non-zero $\phi$ will cause the two angles, $\alpha_3$ and $\alpha_4$, to differ by an amount $\alpha_4 - \alpha_3 = -d \phi$ \cite{Foxen_PRL_2020}. Example experimental values of $\alpha_1 + \alpha_2$ and $\alpha_4 - \alpha_3$ as a function of $d$ are shown in the bottom panel of Fig.~\ref{fig:2s}b. The slopes of both data sets allow $\Delta_+$ and $\phi$ to be extracted, which are seen to be very close to the target condition $2 \Delta_+ = \phi$. Figure~\ref{fig:2s}c shows experimentally calibrated values of $\phi$ across the 20-qubit chain, for a total of 40 disorder instances. It can be seen that $\phi$ falls within the intended range of $\left[ - 0.5 \pi, -1.5 \pi \right]$. Figure~\ref{fig:2s}d shows the calibrated values of $\phi$ when each $\phi_i$ has a target value of $-0.4$. Comparing the measured values of $\phi$ with the target value, we find a small control error of 0.0085 rad for $\phi$.

\subsection{Calibration of $\theta$}

For iSWAP-like gates within the FSIM family, the iSWAP angle $\theta$ can be accurately calibrated by setting $\Delta_- = 0$ and applying FSIM gates in succession, which leads to population transfer between the two qubits (initialized in $\ket{10}$) with a period of $\frac{\pi}{\theta}$. Discrete fourier transform of qubit populations therefore allow $\theta$ to be determined with very high precision \cite{Neill_Nature_2021}. However, CPHASE-like gates typically have small iSWAP angles and such a technique is no longer as effective, since the period for one population transfer can be very long and the calibration data are complicated by noise effects. 

To circumvent such a problem, we have designed a new gate sequence for calibrating small values of $\theta$: Let us consider the composite unitary $\hat{U}_\text{COM} = \hat{U}_\text{FSIM} X_\text{a} Y_\text{b}$ where $X_\text{a}$ and $Y_\text{b}$ are $X$ and $Y$ $\pi$-rotations of $Q_\text{a}$ and $Q_\text{b}$, respectively. $\hat{U}_\text{COM}$ has the following matrix form:
\begin{equation}
\begin{pmatrix}
0 & 0 & 0 & -i \\
0 & -e^{-i  \Delta_{-, \text{off}}} \sin{\theta} & i\cos{\theta} & 0 \\
0 & -i\cos{\theta} & e^{i  \Delta_{-, \text{off}}} \sin{\theta} & 0 \\
i e^{-i \phi} & 0 & 0 & 0
\end{pmatrix}.
\end{equation}

Here we have set $\Delta_-$ and $\Delta_+$ to 0 for simplicity (non-zero values of these phases will not affect the calibration scheme discussed below). With simple algebra, it can be seen that for qubits initialized in the $\ket{10}$ state, the excited state population of $Q_\text{b}$ after applying $\hat{U}_\text{COM}$ $d$ times ($d$ being odd) is bounded by two values: For $\Delta_{-, \text{off}} = 0$, $P_1 = \cos ^ 2 \theta \approx 1$ for small values of $\theta$. For $\Delta_{-, \text{off}} = \frac{\pi}{2}$, $P_1 = \cos ^ 2 (d \theta)$. The difference between these two values is $\cos ^ 2 \theta - \cos ^ 2 (d \theta)$ which is approximately $\sin^2 (d \theta)$ for $\theta \approx 0$. As such, varying $\Delta_{-, \text{off}}$ and measuring the peak-to-peak amplitude of $P_1$ allows determination of $\theta$. Compared to the fourier-transform technique, the method here requires relatively short circuit depth, as a iSWAP angle of 0.02 rad would yield a peak-to-peak amplitude of 0.1 in $P_1$ for $d = 17$, which can be resolved with a relatively small number of single-shot measurements.

The experimental Floquet circuit for calibrating $\theta$ is shown in the upper panel of Fig.~\ref{fig:3s}a. Here, we have injected a variational angle $\psi$ into the single-qubit $\pi$-rotations. Varying $\psi$ effectively changes $\chi$ of the FSIM gate. The experimental calibration data for a given pair of qubits are shown in the bottom panel of Fig.~\ref{fig:3s}a, where we see oscillations of $P_1$ as a function of $\psi$. Fitting the data to a sinusoidal function allows a peak-to-peak amplitude of 0.22 to be determined, which corresponds to a iSWAP angle of $\theta = 0.029$ rad.

The iSWAP angles for all qubit pairs of the 20-qubit chain are shown in Fig.~\ref{fig:3s}b, where we have included data for 40 disorder instances in $\phi$. A small average value of 0.022 rad is found for the qubit chain, with the fluctuation between disorder instances understood as a result of different detunings between the $\ket{01}$ and $\ket{10}$ states during the flux pulses of different gates.


\section{Derivation of effective Hamiltonians}

The bit-string energies shown in the inset to  Fig.~3b are based on effective Hamiltonians $\hat{H}_\text{eff}$ that, in certain limits, approximate the effect of the unitary circuit over two periods, $\hat{U}_F^2 \approx e^{-2i\hat{H}_\text{eff}}$.
Here we derive the relevant $\hat{H}_\text{eff}$ operators for the models in Fig.~3b.

\subsection{Uniform $\phi_i=-0.4$}

For the model with uniform CPHASE angles $\phi_i \equiv \bar{\phi} = -0.4$ and random single-qubit $Z$ rotation angles $h_i\in [-\pi,\pi]$, a period of the time evolution is represented by
\begin{equation}
\hat{U}_F^\prime =  \hat{U}_z[\bar{\phi}, \mathbf{h}]  \hat{U}_x[\pi-2\epsilon]
\end{equation}
with 
\begin{align*}
\hat{U}_x[\theta] 
& = e^{-i\frac{\theta}{2} \sum_n \hat{X}_n}\;,\\
\hat{U}_z[\bar{\phi}, \mathbf{h}] 
& = e^{-i\sum_n (\bar{\phi}/4) \hat{Z}_n \hat{Z}_{n+1} + (h_n/2) \hat{Z}_n} \;.
\end{align*}
We have also defined the detuning $\epsilon = \frac{\pi}{2}(1-g)$;  in the following we take $\epsilon\ll 1$, i.e. $g$ close to 1.
The evolution over two periods is given by
\begin{align}
(\hat{U}_F^\prime)^2 
& = \hat{U}_z[\bar{\phi}, \mathbf{h}] \hat{U}_x[\pi -2\epsilon] \hat{U}_z[\bar{\phi}, \mathbf{h}] \hat{U}_x[\pi-2\epsilon]  \nonumber \\
& = \hat{U}_z[\bar{\phi}, \mathbf{h}] \hat{U}_x[ -2\epsilon] \hat{U}_z[\bar{\phi}, -\mathbf{h}] \hat{U}_x[-2\epsilon] 
\end{align}
where we have used the comutation properties of the perfect $\pi$ pulse $\hat{U}_x[\pi] = \prod_n \hat{X}_n$.
Next, we note that $\hat{U}_z[\bar{\phi}, -\mathbf{h}] = \hat{U}_z[\bar{\phi},0] \hat{U}_z[0,\mathbf{h}]^\dagger$;
acting by conjugation with $\hat{U}_z[0,\mathbf{h}]$ on $\hat{U}_x[-2\epsilon]$ gives
\begin{align}
(\hat{U}_F^\prime)^2 
& = \hat{U}_z[\bar{\phi},0] 
e^{i\epsilon \sum_n \cos(h_n) \hat{X}_n + \sin(h_n)\hat{Y}_n} \hat{U}_z[{\bar\phi},0] \hat{U}_x[-2\epsilon]  \;.
\label{eq:uf2}
\end{align}
The effective Hamiltonian $\hat{H}_\text{eff}$, satisfying $(\hat{U}_F^\prime)^2 \approx e^{-2i\hat{H}_\text{eff}}$, is then given to leading order in $\epsilon$, $|\bar{\phi}/4|\ll 1$ via the Baker-Campbell-Hausdorff (BCH) formula:
\begin{align}
    \hat{H}_\text{eff}
    & = \sum_n \frac{\epsilon}{2} [(1+\cos(h_n))\hat{X}_n + \sin(h_n) \hat{Y}_n] + \nonumber \\
    & \qquad + \frac{\bar{\phi}}{4} \hat{Z}_n \hat{Z}_{n+1}\;.
    \label{eq:Heff1}
\end{align}
Thus, for any bit-string $\mathbf{s}\in \{0,1\}^L$, the energy of the associated computational basis state $\ket{\mathbf{s}} = \ket{s_1}\ket{s_2}\cdots \ket{s_L}$ is
\begin{equation}
E_{\mathbf s} = \bra{\mathbf{s}} \hat{H}_\text{eff} \ket{\mathbf{s}} = \frac{\bar\phi}{4} \sum_n (-1)^{s_n+s_{n+1}} \;,
\label{eq:e_pt}
\end{equation}
which simply counts the number of ``domain walls'' (i.e. bonds where $s_i \neq s_{i+1}$) in $\mathbf{s}$.
Thus the polarized and staggered bit-strings (having 0 and $L-1$ ``domain walls'', respectively) lie near the edges of the spectrum in all realizations.

We note that, strictly speaking, a prethermal DTC requires $\hat{H}_\text{eff}$ to have a symmetry breaking transition at a finite critical temperature $T_c$. 
In this case, ordered initial states at temperatures $T<T_c$ show long-lived oscillations with an amplitude that depends on the equilibrium value of the (symmetry breaking) order parameter at temperature $T$~\cite{Else2017}.
As short-ranged models in one dimension (such as the one under consideration) cannot have order at finite temperature, this model is not prethermal in the sense we just described. 
However, thermal correlation lengths may still exceed the size of the system when the latter is small enough. This allows low-temperature states of $\hat{H}_\text{eff}$ to show long-lived oscillations with a finite amplitude, even if the equilibrium order parameter is asymptotically zero for such states.

\subsection{Disordered $\phi_i \in [-1.5\pi, -0.5\pi]$}

In the MBL DTC drive $\hat{U}_F$ we set the average CPHASE angle to $\bar{\phi} = -\pi$, which (being $\sim 10$ times larger than in the previous case) breaks the final step in the derivation of Eq.~\eqref{eq:Heff1}.
We can however use another approach, valid if $\phi_i = -\pi + \delta\phi_i$, with $|\delta\phi_i|$ sufficiently small. 
In Eq.~\eqref{eq:uf2} we replace $\bar{\phi}$ by $\pi + \boldsymbol{\delta\phi}$, and noting that $\hat{U}_z[\pi,0] = \hat{U}_z[-\pi,0]$ \footnote{This is true up to single-qubit $Z$ rotations on the edge qubits, if the chain has open boundary conditions. These could be cancelled by considering the evolution over 4 periods, with minor changes to the result (a cancellation of terms near the edges). For the sake of simplicity we will neglect this effect here.}
we have
\begin{align}
\hat{U}_F^2 
& = \hat{U}_z[\boldsymbol{\delta\phi},0] 
e^{i\epsilon \sum_n (\cos(h_n)\hat{Y}_n-\sin(h_n)\hat{X}_n)(\hat{Z}_{n-1}+\hat{Z}_{n+1})}
\nonumber \\
& \qquad \times 
\hat{U}_z[\boldsymbol{\delta\phi},0] \hat{U}_x[-2\epsilon] \;.
\end{align}
If $\epsilon$, $|\delta\phi_i| \ll 1$, leading-order BCH yields
\begin{align}
\hat{H}_\text{eff}
& = \sum_n \frac{\epsilon}{2} [\hat{X}_n + \cos(h_n) \hat{Y}_n(\hat{Z}_{n-1} + \hat{Z}_{n+1})]  \nonumber \\
& \ + \frac{\delta\phi_n}{4} \hat{Z}_n \hat{Z}_{n+1} -\frac{\epsilon}{2} \sin(h_n)\hat{X}_n (\hat{Z}_{n-1} + \hat{Z}_{n+1})
\end{align}
The energy of a bit-string state $\ket{\mathbf{s}}$ is 
\begin{equation}
E_{\mathbf s} = \bra{\mathbf s} H_F^{(0)} \ket{\mathbf s} = \sum_n \frac{\delta\phi_n}{4} (-1)^{s_n + s_{n+1}} \;.
\end{equation}
Unlike the result in Eq.~\eqref{eq:e_pt}, this does not single out special bit-strings. More specifically, under disorder averaging all bit-strings have the same energy: $\overline{E_{\mathbf{s}}} = 0$.

In our model, the $|\delta\phi_i|$ angles are not small (they vary in $[-0.5\pi,0.5\pi]$) so all orders in BCH should be included for an accurate result -- the above is only meant to be a qualitative approximation.
Nevertheless, the independence of (disorder-averaged) quantities from the choice of bit-string can be proven exactly for this model. 

All bit-string states are obtained as $\ket{\mathbf{s}} = \hat{X}_\mathbf{s} \ket{\boldsymbol{0}}$, where $\ket{\boldsymbol{0}} = \ket{00\dots 00}$ is the polarized state and $\hat{X}_\mathbf{s} = \prod_{i:s_i=1} \hat{X}_i$ flips the qubits where $s_i = 1$. 
We will show that all bit-string states give rise to the same dynamics as the polarized one, up to a change in the realization of disorder.
Indeed, the change of basis that maps $\ket{\mathbf s}$ to $\ket{\boldsymbol{0}}$ acts on $\hat{U}_F$ as 
\begin{equation}
\hat{X}_\mathbf{s} \hat{U}_F \hat{X}_\mathbf{s}
= \hat{U}_z[\boldsymbol{\phi}^\prime, \mathbf{h}^\prime] \hat{U}_x[\pi g]
\end{equation}
where $\phi_i^\prime = (-1)^{s_i+s_{i+1}} \phi_i$ and $h_i^\prime = (-1)^{s_i} h_i$.
$\boldsymbol{\phi}^\prime$ and $\mathbf{h}^\prime$ almost define another valid realization of disorder, except that wherever $s_i\neq s_{i+1}$, we have $\phi_i^\prime \in [0.5\pi, 1.5\pi]$ (as opposed to $\phi_i \in [-1.5\pi,-0.5\pi]$).
This can be remedied by setting $\phi_i^{\prime\prime} = \phi_i^\prime-2\pi \in [-1.5\pi,-0.5\pi]$, and noting that $e^{-i \frac{\pi}{2}\hat{Z}_i\hat{Z}_{i+1}} \propto e^{-i \frac{\pi}{2}\hat{Z}_i} e^{-i\frac{\pi}{2} \hat{Z}_{i+1}}$, so that the excess $2\pi$ CPHASE angle can be transferred to single-qubit rotations:
$\hat{U}_z[\boldsymbol{\phi}^\prime, \mathbf{h}^\prime] \propto \hat{U}_z[\boldsymbol{\phi}^{\prime\prime}, \mathbf{h}^{\prime\prime}]$, where $h_i^{\prime\prime} = h_i^\prime$ if $s_{i-1}=s_{i+1}$, or $h_i^\prime + \pi$ otherwise.
Thus the dynamics of bit-string $\ket{\mathbf s}$ under disorder realization $(\boldsymbol{\phi},\mathbf{h})$ is mapped to the dynamics of $\ket{\boldsymbol{0}}$ under a different realization $(\boldsymbol{\phi}^{\prime\prime}, \mathbf{h}^{\prime\prime})$. 
Further, the mapping between realizations conserves the (uniform) measure over the intervals $\phi_i\in[-1.5\pi, -0.5\pi]$, $h_i \in [-\pi, \pi]$.
Thus after averaging over disorder, all bit-strings are equivalent to each other.

\section{Echo circuit for noise mitigation}

The ``echo'' circuit $\hat{U}_\text{ECHO}$ used to define the normalization $A_0$ in Fig.~\ref{fig:2}c and Fig.~\ref{fig:4}d of the main text consists of 
$t$ steps of forward time evolution under $\hat{U}_F$ and $t$ steps for ``backward'' time evolution under $\hat{U}_F^\dagger$.
In the absence of noise, the two cancel exactly.
Thus deviations from this outcome quantify the impact of noise.

\subsection{Circuit inversion}

Our device allows the calibration of a continuous family of CPHASE angles on each bond and their use during a single circuit run. Thus it is possible to concatenate the forward and backward time evolutions $\hat{U}_F^t$ and $(\hat{U}_F^\dagger)^t$ directly. However, the two-qubit gates in $\hat{U}_F^\dagger$ would have in general different fidelity than those in $\hat{U}_F$.
As a result, the decoherence during $\hat{U}_\text{ECHO}$ would differ from that during $\hat{U}_F$.

To circumvent this, we effectively invert the circuit $\hat{U}_F$ without changing the two-qubit gates, thus keeping the fidelity unchanged during the backward time evolution. 
The key idea is to apply $X$ rotations on even qubits, $\hat{P}_{\pi,e} \equiv \prod_{n=1}^{L/2} \hat{X}_{2n}$, before and after each period of the circuit that is to be inverted.
Indeed conjugating $\hat{U}_F$ by $\hat{P}_{\pi,e}$ changes the sign of all $\phi_n$ CPHASE angles.
It also changes the sign of single-qubit $Z$ rotation angles $h_n$ on even sites. 
The sign of remaining $h_n$ fields and of $g$, as well as the relative order of the $X$ and $Z$ parts of the drive, can be changed at the level of single-qubit gates, with minor effects on fidelity.

In practice, after $t$ cycles of $\hat{U}_F$, we apply the single-qubit rotations $\hat{P}_{\pi,e}$ only once, and then switch to a unitary $\hat{V}_F$ which has the same 2-qubit gates as $\hat{U}_F$ but different single-qubit gates chosen so that $\hat{P}_{\pi,e} \hat{V}_F \hat{P}_{\pi,e} = \hat{U}_F^\dagger$  (as explained above).
Finally we measure in the computational basis and flip the logical value of all bits at even positions (this is equivalent to acting with $\hat{P}_{\pi,e}$ at the final time but avoids any fidelity losses). 
This way, we manage to effectively invert the circuit without altering two-qubit gate fidelities.


\subsection{Measuring the effect of decoherence}

Let us model noise as a single-qubit channel 
\begin{equation}
\mathcal{E}_p = \left( 1- \frac{4p}{3} \right)\mathcal{I} + \frac{4p}{3} \mathcal{D}
\end{equation}
where $\mathcal{I}$ is the identity channel ($\mathcal{I}(\hat{\rho}) = \hat{\rho}$), $\mathcal{D}$ is the single-qubit erasure channel ($\mathcal{D}(\hat{\rho})\propto I$), and $p$ denotes the error rate.
We assume the noise acts symmetrically before and after each iteration of the unitary circuit (while realistically noise acts \emph{during} the entire evolution, this is a good approximation as long as $p\ll 1$).
Then the time evolution of the system over a period is described by a quantum channel
\begin{equation}
\hat{\rho} \mapsto \mathcal{E}_{p/2}^{\otimes L} \circ \mathcal{U}_F \circ \mathcal{E}_{p/2}^{\otimes L} (\hat{\rho}) \equiv \Phi(\hat{\rho})
\end{equation}
where $\mathcal{U}_F(\hat{\rho}) = \hat{U}_F \hat{\rho} \hat{U}_F^\dagger$.
Similarly the inverted time evolution is given by 
\begin{equation}
\hat{\rho} \mapsto \mathcal{E}_{p/2}^{\otimes L} \circ \mathcal{U}^\dagger_F \circ \mathcal{E}_{p/2}^{\otimes L} (\hat{\rho}) \equiv \Phi^\dagger (\hat{\rho})
\end{equation}
where $\mathcal{U}^\dagger_F(\hat{\rho}) = \hat{U}_F^\dagger \hat{\rho} \hat{U}_F$, and the last equality holds because $\mathcal{E}_p$ is self-adjoint.
The entire echo circuit sequence is thus described by the channel $(\Phi^\dagger)^t \circ \Phi^t$.
The expectation value of $\hat{Z}_i$ after the circuit is given by 
\begin{align}
\langle \hat{Z}_i\rangle^\text{echo}_{\mathbf{s}}
& \equiv \text{Tr} \left( \hat{Z}_i (\Phi^\dagger)^t \circ \Phi^t (\ket{\mathbf{s}}\bra{\mathbf{s}}) \right)
\end{align}
The temporal autocorrelator between $\hat{Z}_i$ before and after the echo circuit is simply $(-1)^{s_i} \langle \hat{Z}_i \rangle^\text{echo}_{\mathbf{s}}$. 
Averaging over all bit-strings yields
\begin{align}
A_0^2 
& \equiv \frac{1}{2^L} \sum_{\mathbf s} (-1)^{s_i} \langle \hat{Z}_i\rangle^\text{echo}_{\mathbf{s}} \nonumber \\
& = \frac{1}{2^L} \sum_{\mathbf s} \text{Tr} \left[ \hat{Z}_i (\Phi^\dagger)^t \circ \Phi^t ( \hat{Z}_i \ket{\mathbf{s}}\bra{\mathbf{s}}) \right] \nonumber \\
& = \frac{1}{2^L} \text{Tr}\left[ (\Phi^t(\hat{Z}_i))^2 \right]
 = \|\Phi^t(\hat{Z}_i)\|^2 / \| \hat{Z}_i \|^2 \;,
\end{align}
where we have used the definition of adjoint channel, $(\hat{V}, \Phi(\hat{W})) = (\Phi^\dagger(\hat{V}), \hat{W})$, relative to the trace inner product $(\hat{V}, \hat{W}) = \text{Tr}(\hat{V}^\dagger \hat{W})$.
Thus from the protocol outlined above one extracts the decay of operator norm $\| \hat{Z}_i(t) \| \sim A_0 \| \hat{Z}_i(0)\|$ which is the leading effect of decoherence.
The ratio $A/A_0$ in Fig.~\ref{fig:2s}d thus gives the overlap between \emph{normalized} operators,
\begin{equation}
    \frac{A}{A_0} = \left( \frac{\hat{Z}_i(0)}{\|\hat{Z}_i(0)\|} \bigg| \frac{\hat{Z}_i(t)}{\|\hat{Z}_i(t)\|} \right) \;.
\end{equation}

\section{Spectral averages via quantum typicality}

Quantum typicality\cite{Popescu2006, Goldstein2006, Gemmer2009} states that, for any observable $\hat{O}$ in a Hilbert space of dimension $2^L$, the expectation value $\langle \hat{O}\rangle_\psi$ on a \emph{random state} $\psi$ sampled from the unitarily invariant (Haar) measure obeys these statistical properties:
\begin{align}
\mathbb{E}_\psi \langle \hat{O}\rangle_\psi
& = \langle \hat{O} \rangle_{\infty} \label{eq:qtyp_mean} \\
\text{Var}_\psi \langle \hat{O}\rangle_\psi
& = \frac{1}{2^L+1} \left( \langle \hat{O}^2 \rangle_\infty - \langle \hat{O} \rangle_\infty^2 \right) \label{eq:qtyp_var}
\end{align}
where $\langle \hat{O} \rangle_\infty \equiv 2^{-L} \text{Tr} (\hat{O})$ denotes the expectation value on the infinite-temperature state.
Thus for large $L$, the matrix element $\langle\hat{O}\rangle_\psi$ is distributed as a Gaussian centered at the infinite-temperature value with an extremely narrow standard deviation, $\simeq 2^{-L/2}$, which enables the measurement of spectrally-averaged quantities with exponentially high accuracy from a single pure state.

\subsection{Scrambling circuit and approach to typicality}

\begin{figure}
    \centering
    \includegraphics[width=\columnwidth]{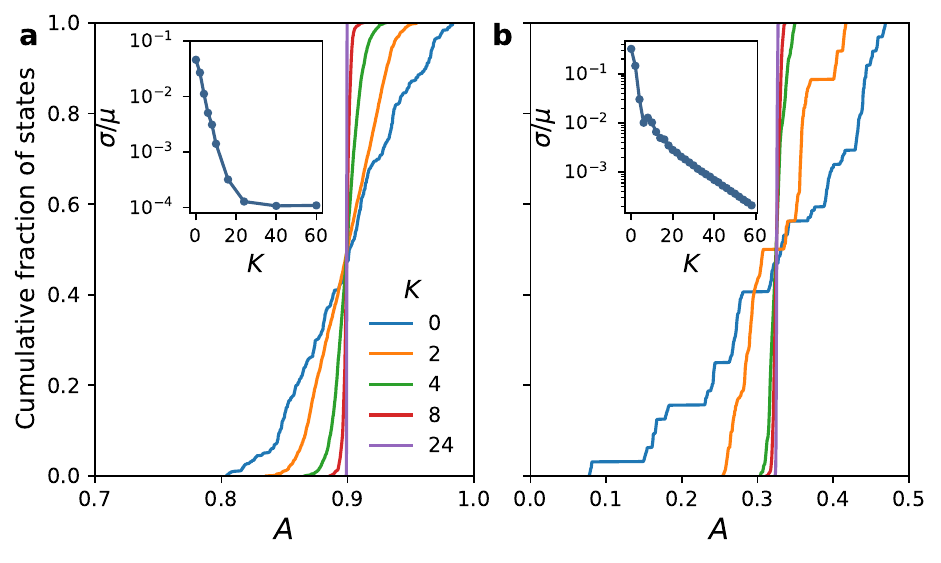}
    \caption{{\bf Simulation of quantum typicality protocol.} 
    {\bf a}, Cumulative distribution of autocorrelators $A$ from a set of 2000 bit-string states pre-processed by a depth-$K$ random circuit $\hat{U}_S$ as described in the text, for variable $K$.
    We set $g=0.94$ (MBL DTC phase).
    The realization of disorder is fixed and $A$ is computed at time $t=30$ on qubit $Q_{11}$ in a chain of $L=20$ qubits.
    Inset: relative standard deviation $\sigma/\mu$ decreases exponentially in $K$ and approaches the random-state variance ($<2^{-L/2}$) after depth $K\simeq L$.
    {\bf b}, Same plot for noisy simulations (depolarizing noise, error rate $p=0.5\%$ per 2-qubit gate, exact density matrix simulations) of qubit $Q_7$ in a chain of $L=12$ qubits, where we include all 4096 bit-string states. 
    Inset: relative standard deviation $\sigma/\mu$. $\sigma$ decreases indefinitely with $K$ due to decoherence, while $\mu$ is not affected.
    }
    \label{fig:typ}
\end{figure}

In the main text we report data on spectrally-averaged autocorrelators $\langle \hat{Z}_i(0) \hat{Z}_i(t) \rangle_\infty$ obtained with a method based on the idea above, i.e by evaluating $\langle \hat{Z}_i(0) \hat{Z}_i(t) \rangle_\psi$ on a state $\ket{\psi}$ which is close to typical random states in the Hilbert space.
In order to prepare a random state $\ket{\psi}$, we start with a bit-string state and evolve it via a scrambling circuit $\hat{U}_S$, as also proposed in Ref.~\cite{Pal2021}.
This consists of $K$ layers of CZ gates (CPHASE with angle $\phi = \pi$) and random single-qubit gates (rotations around a random direction on the $XY$ plane by an angle $\theta$ sampled uniformly in $[0.4\pi, 0.6\pi]$).
The single-qubit gates vary randomly in both space and time, so that $\hat{U}_S$ is not a Floquet circuit. 
After a number of layers $K = O(L)$ (we neglect decoherence for now), the prepared state $\ket{\psi} = \hat{U}_S \ket{\mathbf{s}}$ behaves like typical random vectors in the Hilbert space, so that $\bra{\psi} \hat{O} \ket{\psi} = \langle \hat{O} \rangle_\infty + \delta$, where the error $\delta$ (a random variable dependent on the choice of bit-string $\mathbf s$ and of scrambling circuit $\hat{U}_S$) has zero mean and variance $\sim 2^{-\min(L,cK)}$ for some constant $c>0$, i.e., the variance shrinks with increasing $K$ as the state becomes gradually more random, until it saturates to the quantum typicality result Eq.~\eqref{eq:qtyp_var}.
In Fig.~\ref{fig:typ}a we show the results of exact numerical simulations that confirm this picture.
For this family of random circuits, we find $c \simeq 0.36$ (from a fit to the data in the inset to Fig.~\ref{fig:typ}a).

\subsection{Ancilla protocol}

To measure two-time correlators $\bra{\psi} \hat{Z}_i(0) \hat{Z}_i(t) \ket{\psi}$ in the ``pseudorandom states'' $\ket{\psi}$ defined above, we use an interferometric protocol similar to the one employed in Ref.~\cite{Mi_OTOC_2021}.
We introduce an ancilla qubit initialized in state $\ket{+X} = (\ket{0} + \ket{1})/\sqrt{2}$ alongside the system qubits $Q_1$, $\dots Q_L$ which are initialized in a bit-string state $\ket{\mathbf{s}}$.
We evolve the system qubits with the scrambling circuit $\hat{U}_S$ for depth $K$, obtaining a joint state $ \ket{\psi}_\text{sys} \ket{+X}_\text{a}$.
Then we apply a CZ gate between the ancilla and system qubit $i$, so that the state ``branches'' into the superposition
\begin{equation}
\frac{1}{\sqrt 2} \left( \ket{\psi}_\text{sys} \ket{0}_\text{a} + \hat{Z}_i \ket{\psi}_\text{sys} \ket{1}_\text{a}\right)
\end{equation}
We then evolve the system under the Floquet drive $\hat{U}_F$ for $t$ periods and again apply a CZ between the ancilla and system qubit $i$, which gives
\begin{equation}
\frac{1}{\sqrt 2} \left( \hat{U}_F^t \ket{\psi}_\text{sys} \ket{0}_\text{a} 
+ \hat{Z}_i \hat{U}_F^t \hat{Z}_i \ket{\psi}_\text{sys} \ket{1}_\text{a}\right)
\end{equation}
Finally, we measure the ancilla in the $X$ basis. The expectation value of the measurement is
\begin{align}
\langle \hat{X}_\text{a} \rangle
& = \frac{1}{2} \bra{\psi} \hat{U}_F^{-t} \hat{Z}_i \hat{U}_F^t \hat{Z}_i \ket{\psi}+ \text{c.c.} \nonumber \\
& = \text{Re} \bra{\psi}\hat{Z}_i(t) \hat{Z}_i(0) \ket{\psi} \nonumber \\
& \simeq \langle \hat{Z}_i(0) \hat{Z}_i(t) \rangle_\infty
\label{eq:ancilla_result}
\end{align}
where the last line follows from quantum typicality if $\ket{\psi}$ is random.
On a sufficiently large system, and for large enough $K$ (number of scrambling cycles), this protocol gives the spectrum-averaged temporal autocorrelator from a single measurement.

\subsection{Effect of noise during the scrambling circuit}

The above discussion neglects decoherence and treats the states during the protocol as pure. Since $K$ must grow with $L$ for the protocol to succeed, it is especially important to understand whether noise during the random state preparation process has a negative impact on the result.

One can repeat the previous discussion with quantum channels instead of unitary operators: the system starts in pure state $\ket{\mathbf s}\! \bra{\mathbf s}_\text{sys} \ket{+X}\!\bra{+X}_\text{a}$ and evolves under the scrambling dynamics into $\hat{\rho}_{\text{sys}} \ket{+X}\!\bra{+X}_\text{a}$, where $\hat{\rho} = \Phi_S(\ket{\mathbf s}\!\bra{\mathbf s})$ and $\Phi_S$ is a quantum channel representing the noisy version of the scrambling circuit $\hat{U}_S$
(we neglect decoherence on the ancilla qubit). 
The protocol then proceeds analogously to the noiseless case and yields the final state
\begin{align}
\hat{\rho}_\text{sys,a}^\prime
= \frac{1}{2} \big[ 
& \Phi^t (\hat{\rho})_\text{sys} \ket{0}\!\bra{0}_\text{a}
+ (\hat{Z}_i \Phi^t (\hat{Z}_i \hat{\rho}))_\text{sys} \ket{1}\!\bra{0}_\text{a} \nonumber \\
& + (\Phi^t (\hat{\rho} \hat{Z}_i)\hat{Z}_i)_\text{sys} \ket{0}\!\bra{1}_\text{a}  \nonumber \\
& + (\hat{Z}_i \Phi^t ( \hat{Z}_i \hat{\rho} \hat{Z}_i ) \hat{Z}_i)_\text{sys} \ket{1}\!\bra{1}_\text{a} 
\big]
\end{align}
where $\Phi$ is the noisy version of the Floquet evolution $\hat{U}_F$.
The expectation of $\hat{X}_\text{a}$ on this state is
\begin{align}
\langle \hat{X}_\text{a} \rangle
& = \frac{1}{2} \text{Tr} \left[ \hat{Z}_i \Phi^t (\hat{Z}_i \hat{\rho}) + \Phi^t (\hat{\rho} \hat{Z}_i) \hat{Z}_i \right] \nonumber \\
& = \frac{1}{2} \text{Tr}[ \{ \hat{Z}_i(t),  \hat{Z}_i(0) \} \hat{\rho}]
\label{eq:noisy_stateprep}
\end{align}
where we have defined $\hat{Z}_i(t) = (\Phi^\dagger)^t [\hat{Z}_i]$ as the Heisenberg-picture evolution of $\hat{Z}_i$ with noise.

To see that Eq.~\eqref{eq:noisy_stateprep} approximates the infinite-temperature expectation value $\langle \hat{Z}_i(0) \hat{Z}_i(t) \rangle_\infty$, we observe that under a random unitary circuit, noise can be approximated by a global depolarizing channel\cite{Arute2019}:
$\Phi_S(\ket{\mathbf{s}} \bra{\mathbf{s}} ) \approx f^K \hat{U}_S \ket{\mathbf s}\bra{\mathbf s} \hat{U}_S^\dagger + (1-f^K) \hat{I}/2^L$, i.e. a sum of the ideal evolution under $\hat{U}_S$ and the fully mixed state $\hat{I}/2^L$ ($\hat{I}$ is the identity matrix), parametrized by a fidelity $f<1$. 
However both the ideal scrambled state $\hat{U}_S \ket{\mathbf s}$ and the fully-mixed state $\hat{I}/2^L$ accurately reproduce the infinite-temperature expectation value.
Thus decoherence during the random state preparation process may in fact be helpful, rather than harmful (as long as the circuit $\hat{U}_S$ is temporally random, so as to avoid any nontrivial steady states).
This is confirmed by the results of exact density matrix simulations of $L=12$ qubits in the presence of depolarizing noise, in Fig.~\ref{fig:typ}b.
The variance between bit-string states falls exponentially in $K$ (depth of $\hat{U}_S$) even below the ideal quantum typicality limit of Eq.~\eqref{eq:qtyp_var}. 
The subsequent decay is purely due to decoherence: the scrambled state $\Phi_S(\hat{\rho})$ asymptotically approaches the fully mixed state $\hat{I}/2^L$ as the noisy circuit is made deeper.

\section{Numerical results on spin glass order parameter}

\begin{figure}
    \centering
    \includegraphics[width=\columnwidth]{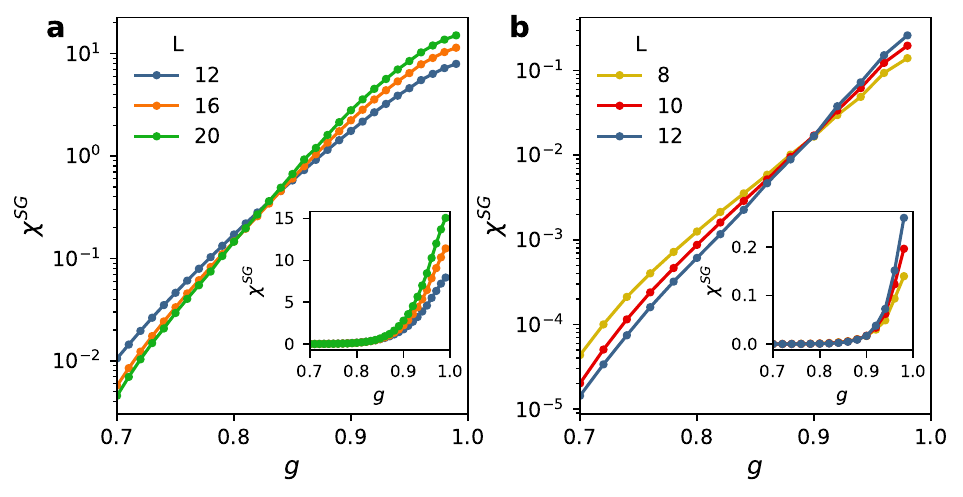}
    \caption{{\bf Numerical simulations of spin glass order parameter.} {\bf a}, Ideal (noiseless) dynamics. $\chi^{SG}$ averaged over even times between $t=50$ and $t=60$, over initial bit-string states and over realizations of disorder. At least 4000 realizations are averaged at sizes $L=12$ and 16, at least 300 at $L=20$. Inset: same data on a linear scale.
    {\bf b}, Noisy dynamics (exact density matrix simulations). Noise is modeled by single-qubit depolarizing channels with Pauli error rate $p=0.5\%$ after each 2-qubit gate. $\chi^{SG}$ is averaged over even times between $t=50$ and $t=60$, over initial bit-string states, and over realizations of disorder. At least 1000 realizations are used at sizes $L=8$ and 10, at least 100 at $L=12$. Inset: same data on a linear scale.}
    \label{fig:chi_sg}
\end{figure}

\begin{figure}[t!]
    \centering
    \includegraphics[width=\columnwidth]{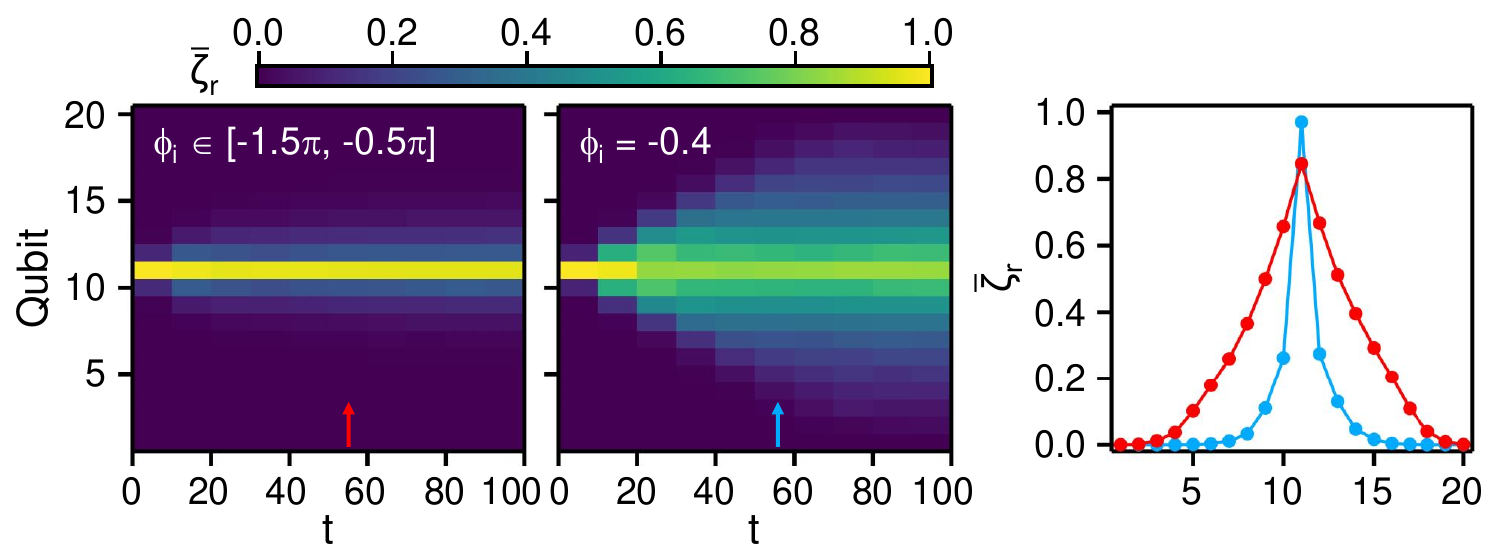}
    \caption{\textbf{Numerical simulations of correlation measurements.} Noiseless simulation of the experiment in Fig.~\ref{fig:3}d of the main text. Here we simulate the fractional change in $\braket{\hat{Z}(t)}$, $\overline{\zeta}_r$ (see definition in main text), due to a single bit-flip at $Q_{11}$ in initial condition. The simulation is averaged over 1000 disorder instances for both $\phi_i \in [-1.5 \pi, -0.5 \pi]$ and $\phi_i = -0.4$.}
    \label{fig:s6}
\end{figure}

Here we show results of numerical simulations of the spin glass order parameter used to perform a finite-size analysis of the phases in the main text.
We define the order parameter as \begin{equation}
\chi^{SG} = \frac{1}{L-2} \sum_{i\neq j}{}^\prime \langle \hat{Z}_i \hat{Z}_j \rangle^2
\end{equation}
where the primed sum excludes the edges (qubits $Q_1$ and $Q_L$) in order to remove the effects of edge modes from bulk physics.
In a phase with glassy order all the expectation values $\langle \hat{Z}_i \hat{Z}_j \rangle$ are finite and $\chi^{SG}$ is extensive ($ \sim L$).
Otherwise, all expectation values asymptotically vanish and $\chi^{SG}/L \to 0$.

In Fig.~\ref{fig:chi_sg}a we show results of numerical simulations of $\chi^{SG}$ in the absence of noise, at times $t$ between 50 and 60 cycles, as the length of the qubit chain is scaled from $L=12$ to $L=20$.
A finite-size crossing is visible near $g \simeq 0.83$, separating a side of parameter space (at larger $g$) where $\chi^{SG}$ grows with $L$, indicative of the MBL-DTC phase, from one (at smaller $g$) where $\chi^{SG}$ decreases with $L$, indicative of a thermalizing phase. 
We also note that the finite-size crossing in these data slowly drifts towards higher $g$ as $t$ increases (not shown), as expected from slow thermalization on the ergodic side near the transition.

Repeating the same analysis in the presence of noise yields the data in Fig.~\ref{fig:chi_sg}b.
We model noise as a single-qubit depolarizing channel with Pauli error rate $p = 0.5\%$ acting on both qubits after each 2-qubit gate.
Simulations are carried out by exact evolution of the density matrix, which is memory-intensive and limits the available system sizes to $L\leq 12$ within reasonable computational resources. 
(We use this method rather than quantum trajectories\cite{Dalibard1992} because the latter method is impractical for this calculation: as $\chi^{SG}$ is a nonlinear function of the state, each expectation $\langle \hat{Z}_i \hat{Z}_j\rangle$ must be averaged over trajectories separately for each disorder realization).
We still find a finite-size crossing, though at a considerably higher value of $g \simeq 0.90$.
We note that the noisy simulations in Fig.~\ref{fig:chi_sg}b do not include the effects of read-out error and that the depolarizing noise model is not guaranteed to be a quantitatively accurate approximation in these structured circuits.
Even so, the experimental estimate of the phase transition point $0.84\lesssim g_c \lesssim 0.88$ is reasonably close to the numerical ones, $g_c\simeq 0.83$ (without noise) and $g_c \simeq 0.90$ (with noise).

\section{Numerical Results of Correlation Measurements}

The noiseless simulation of the experiment in Fig.~\ref{fig:3}d of the main text is shown in Fig.~\ref{fig:s6}. Here we generate a total of 1000 disorder instances for both $\phi_i \in [-1.5 \pi, -0.5 \pi]$ and $\phi_i = -0.4$. The values of $\braket{\hat{Z} (t)}$ are simulated with the two initial conditions $\ket{00000000000000000000}$ ($\zeta_1$) and $\ket{00000000001000000000}$ ($\zeta_2$), and the ratio $\overline{\zeta}_\text{r}$ is computed using the same method as the main text.

It is seen that the ratio $\overline{\zeta}_\text{r}$ from the noise simulation is quite similar to the experimentally measured values, despite no active error-mitigation for this particular quantity. This is likely attributed to the fact that decoherence introduces a damping factor that is approximately the same for both the nominator ($|\zeta_1 - \zeta_2|$) and denominator ($|\zeta_1| + |\zeta_2|$) used to compute $\overline{\zeta}_\text{r}$. Consequently, their effects are canceled out after dividing the two quantities.

\end{document}